\documentclass[prx,reprint,amsmath,amssymb,aps,floatfix,dvipsnames,longbibliography]{revtex4-2}

\usepackage[normalem]{ulem}
\usepackage{cancel}
\usepackage{subfiles}
\usepackage{natbib}
\usepackage{graphicx}
\usepackage{dcolumn}
\usepackage{bm}
\usepackage{xcolor}

\newcommand{\change}[1]{\textcolor{black}{\textrm{#1}}}

\usepackage[version=4]{mhchem}

\begin{document}

\title{Quantum symmetrization transition in superconducting sulfur hydride from quantum Monte Carlo and path integral molecular dynamics}

\author{Romain Taureau\textsuperscript{1}}
\author{Marco Cherubini\textsuperscript{1}}
\author{Tommaso Morres\textsuperscript{2}}
\author{Michele Casula\textsuperscript{1}}
\email{michele.casula@sorbonne-universite.fr}
\affiliation{
\textsuperscript{1}Sorbonne Universit\'e, Institut de min\'eralogie, de physique des mat\'eriaux et de cosmochimie (IMPMC),  CNRS UMR 7590, MNHM, 4 Place Jussieu, 75005 Paris.\\
\textsuperscript{2}European Center for Theoretical Studies in Nuclear
Physics and Related Areas, Fondazione Bruno Kessler, Strada delle
Tabarelle 286, Trento, 38123, Italy.
}

\date{\today}

\begin{abstract}
  We study the structural phase transition,  originally associated
  with the highest superconducting critical temperature $T_c$ measured
  in high-pressure sulfur hydride. A quantitative description of its
  pressure dependence has been elusive for any \emph{ab initio} theory
  attempted so far, raising questions on the actual mechanism leading to the maximum of $T_c$.
Here, we estimate the critical pressure of the hydrogen bond symmetrization in the Im$\bar{3}$m structure, 
by combining density functional theory and quantum Monte Carlo simulations for
electrons with path integral molecular dynamics for quantum nuclei.
We find that the $T_c$ maximum corresponds to pressures where local dipole moments dynamically form
  on the hydrogen sites, as precursors of the ferroelectric
  Im$\bar{3}$m-R3m transition, happening at lower pressures.
For comparison, we also apply the self-consistent harmonic approximation,
whose ferroelectric critical pressure lies in between the ferroelectric transition estimated by path integral
  molecular dynamics and the local dipole formation. Nuclear quantum effects 
play a major role in a significant reduction ($\approx$ 50 GPa) of the classical
ferroelectric transition pressure at 200K and in a
large isotope shift ($\approx$ 25 GPa) upon hydrogen-to-deuterium
substitution of the local dipole formation pressure, in agreement
with the corresponding change in the $T_c$ maximum location.
\end{abstract}

\maketitle

\section{Introduction}\label{sec1}

Since its discovery in 1911 \cite{Onnes_supra}, superconductivity has been 
one of the most investigated topics  
in both theoretical and experimental physics.
While it was discovered that almost every conductor could reach 
zero resistance
at low-enough temperatures (T $<$ 10 K) \cite{Tresca2022}, the quest
for higher critical temperature ($T_c$) superconductors became the new
challenge. Until recently, cuprates were leading the race with a $T_c$
as large as 133 K for Hg-Ba-Ca-Cu-O systems \cite{Schilling_1993},
although the pairing mechanism in these materials is considered
unconventional and it is not explained by the standard
Bardeen–Cooper–Schrieffer (BCS) theory \cite{Bardeen_1957}.

In 2015, the discovery of conventional superconductivity in $\ce{H_3S}$ with a maximum $T_c$ of 203 K reached at a pressure $P_c$ as high as 150 GPa \cite{Drozdov_2015} paved the way to a new era of high-$T_c$ materials. Indeed, hydrogen (H) -based systems are nowadays the most promising candidates to achieve room-temperature superconductivity.
As a matter of fact, in 2019, the same team that discovered $\ce{H_3S}$ claimed to have measured an even higher $T_c$ in $\ce{LaH_{10}}$, 
superconducting
already at 250 K \cite{Drozdov_2019}, later followed by a similar
discovery in the yttrium hydride \cite{Kong2021}. In a rush towards
room-temperature superconductivity, more recent claims of $T_c$ larger
than the one found in $\ce{LaH_{10}}$ did not meet the consensus of
the whole community
\cite{Service_2020,Boeri_2023}.
The main issue of these materials is the extreme pressure conditions, 
usually
larger than 150 GPa, needed to obtain the high-$T_c$ superconducting phase. 
Indeed,
while all the binary candidates involving hydrogen were theoretically investigated, none of them seems to sufficiently decrease the pressure of the superconducting state. Eyes are now turned towards ternary materials \cite{DiCataldo2022}.

\begin{figure}[htb!]
    \centering
    \includegraphics[width=0.99
    \linewidth,trim={0cm 0cm 0cm 0cm},clip]{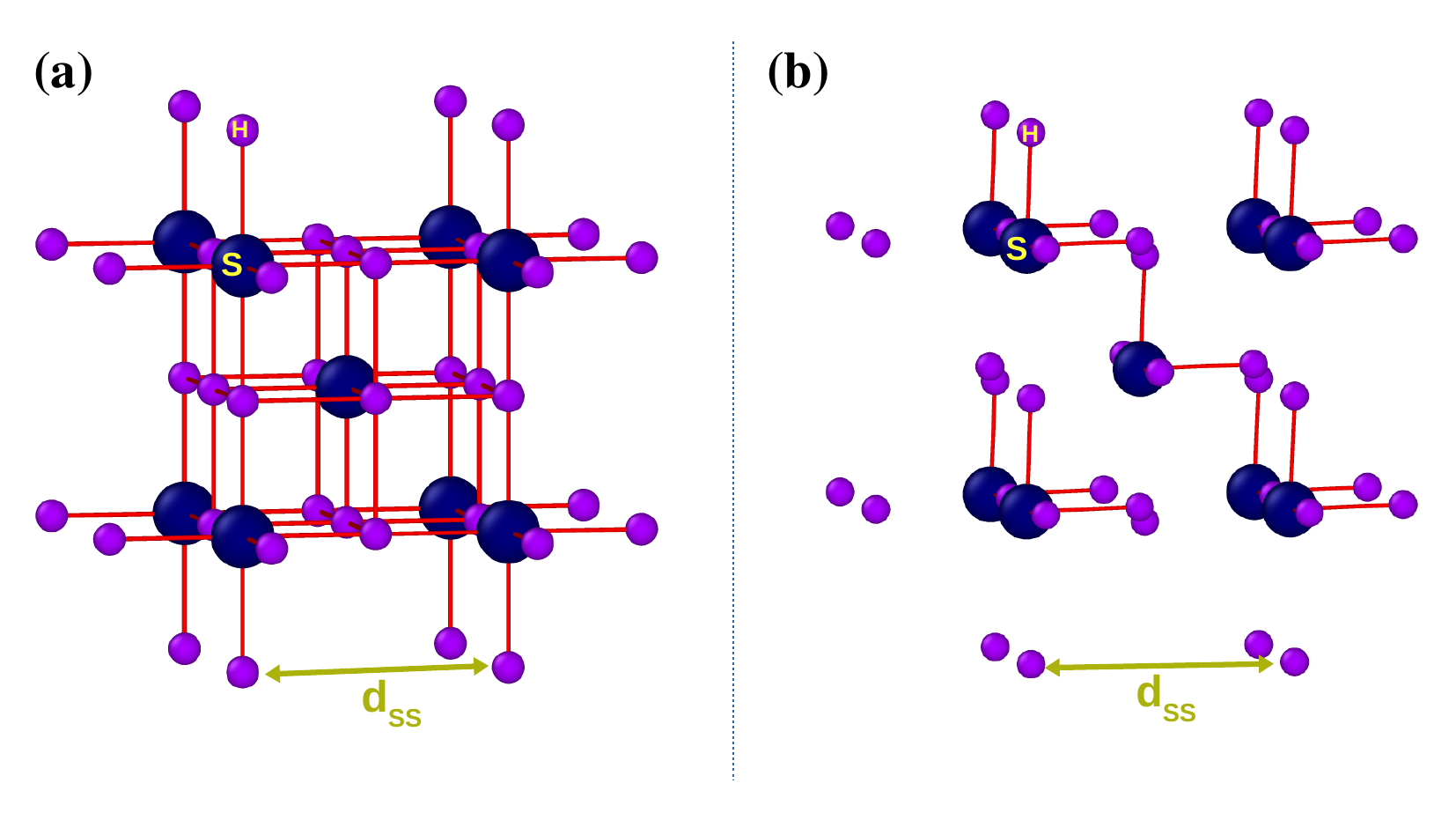}
    \caption{(a) Crystal structure of the Im$\bar{3}$m symmetric phase
      of $\ce{H_3S}$ (smaller volume, higher-pressure phase). (b)
    The R$3$m asymmetric phase 
    (larger volume, lower-pressure phase). d$_{SS}$ is the lattice parameter of the bcc crystal.
    }
    \label{fig:geometry}
\end{figure}

In this work, we focus on the prototypical case of $\ce{H_3S}$ and we study its structural phase transition generally associated with the maximum of 
the  
superconducting critical temperature, located at around 150 GPa \cite{Einaga_2016,mozaffari_2019,Minkov_2020,Osmond_2022}. 
According to x-ray diffraction data \cite{goncharov_2017},   
at lower pressures the sulfur (S) sites are arranged in a geometry that is compatible with
the trigonal R3m 
symmetry
(Fig.~\ref{fig:geometry}(b)) and, upon compression, 
the system
undergoes a phase transition towards a body-centered-cubic (bcc) Im$\bar{3}$m 
structure
(Fig.~\ref{fig:geometry}(a)).

After the first theoretical prediction of high-$T_c$ superconductivity in H$_3$S \cite{duan2014pressure}, several works
tried to explain 
the origin of the maximum of $T_c$ found in experiments as a function of pressure. 
Even if
the magnitude of the calculated $T_c$ is right, confirming the BCS origin of the superconducting state, a quantitative disagreement between various theoretical approaches was found, with estimated $T_c$ values fluctuating over a 50 K range for the high-pressure phase \cite{Sano_2016}. Moreover, theoretical studies more oriented to understand the underlying structural properties of H$_3$S, revealed a significant disagreement in the transition pressures between the predicted phases. In those works \cite{Errea_2016,Bianco_2018}, the structural phase transition is explained by a quantum proton symmetrization from the R3m phase, with displaced protons, to the Im$\bar{3}$m one, where every hydrogen 
lies in the midpoint of the two neighboring sulfur 
atoms (S-S midpoint).
\change{This is also called ferroelectric transition, because
  the hydrogen atoms displaced from the S-S midpoint
  lead in the R3m phase to a long-range order of local
  dipole moments, created by the H-S bond asymmetry.}
In 
that
context, the shuttling mode of 
hydrogen atoms, namely their 
vibrational mode along the direction linking 
two neighboring sulfur atoms, 
was
thoroughly investigated. 
The phase transition 
was then
identified by looking at the dynamical instability of the symmetric Im$\bar{3}$m phase when the pressure is lowered and the shuttling mode softens.
On general grounds,
this
reflects the sudden transformation of the free energy profile, leading to a sign change of its curvature across the transition between two different crystal structures, one with lower symmetry than the other.

These findings were obtained by solving the nuclear Hamiltonian within the Stochastic Self Consistent Harmonic Approximation (SSCHA) \cite{Monacelli2021,Errea_2013,Bianco_2017}, which 
has proven
to be one of the best approximated theories to deal with nuclear quantum effects (NQE). Within this framework, 
the electronic part 
was
solved 
by
Density Functional Theory (DFT) using different parametrizations for the exchange-correlation functional, like the Perdew-Burke-Ernzerhof (PBE) \cite{Perdew1996} and the Becke-Lee-Yang-Parr (BLYP) \cite{Becke_1988,Lee_1988} ones. Independently of the DFT functional used, 
a sizable underestimation of the experimental critical pressure $P_c$ by $\approx$ 40 GPa 
was always
observed, leaving open the question about the origin of 
this
mismatch, and whether this should be attributed to the electronic or to the nuclear 
components.

Here, we go beyond the previous state-of-the-art calculations by treating the electronic 
problem
not only at the DFT-BLYP, but also at Quantum Monte Carlo (QMC) level,
which provides a
benchmark for the 
DFT methods.
QMC is known to 
yield
very accurate total energies in both molecules and solids
\cite{lester_2009,saritas2017investigation,raghav2023toward}, thanks
to its stochastic Green's function algorithms
\cite{Foulkes2001,Wagner2016}, such as the lattice regularized
diffusion Monte Carlo \cite{Casula_lrdmc}, projecting any initial
trial wavefunction towards the ground state of the system within the
fixed node approximation. Moreover, we solve the nuclear Hamiltonian
by using Path Integral Molecular Dynamics (PIMD), which is in
principle exact, outperforming any other approximation for the nuclear
degrees of freedom. Then,
\change{we analyze the resulting phase diagram}
by looking \change{at the
  ferroelectric order parameter,} 
  at the hydrogen/deuterium 
density,
focusing on its 
transformation 
from the unimodal to bimodal distribution, and \change{finally} at its quantum fluctuations, 
detecting
when 
\change{the associated local polarization}
freezes
in a displaced geometry.

In this work, 
we 
have been
able to track the evolution of the mode distribution 
with a high resolution in volume (and pressure),
thanks to a three-dimensional (3D) model of the shuttling mode.
The reliability of 
our
model
has been
benchmarked using \emph{ab initio} PIMD simulations 
with BLYP electrons,
across
the 
\change{local moment formation}. 
The advantage of the 3D model is that its potential energy surface (PES) can still be derived by much more
expensive,
although more 
accurate,
QMC calculations, allowing us to check the impact of the electronic
description on the
\change{occurrence of a local polarization}.

In the model we developed, all hydrogen atoms in the system are 
allowed
to move in the same way.
\change{However, only the spatial degrees of freedom of a single H site are retained.}
This feature induces some limitations,
such as the lack of spatially disordered H configurations, \change{and
  of correlations beyond a single-site description}. 
In spite of this,
we can 
accurately
describe 
the \change{local} path from the symmetric
\change{proton arrangement}
to the 
asymmetric
\change{one, by detecting the local 
moment
  formation in the system, related to the shuttling mode softening.}
We have finally
performed both 
SSCHA and PIMD simulations of the 3D model to 
investigate
how NQE treated at different levels of approximation 
affect the final outcome.

\section{Results}\label{sec2}

\subsection{Harmonic and anharmonic phonons}

At high pressure, above 150 GPa, the $\ce{H_3S}$ crystal is expected to be in the cubic Im$\bar{3}$m symmetric phase (Fig.~\ref{fig:geometry}(a)), where every hydrogen atom 
sits on
the midpoint of 
two neighboring sulfur atoms. Upon 
pressure release,
the lattice 
undergoes
a trigonal distortion and the hydrogen atoms 
leave
the aforementioned midpoint to 
move
closer to one of the two flanking sulfur atoms, leading to the R3m asymmetric phase, depicted in Fig.~\ref{fig:geometry}(b). In our 
description,
we introduce a 
simplification by neglecting the trigonal distortion, which is however very weak ($<$0.6°) \cite{Bianco_2018}. Thus, the R3m phase considered here 
differs from
the Im$\bar{3}$m one just by the 
hydrogen
positions.

\begin{figure*}[htb!]%
\centering
\includegraphics[width=1.\textwidth]{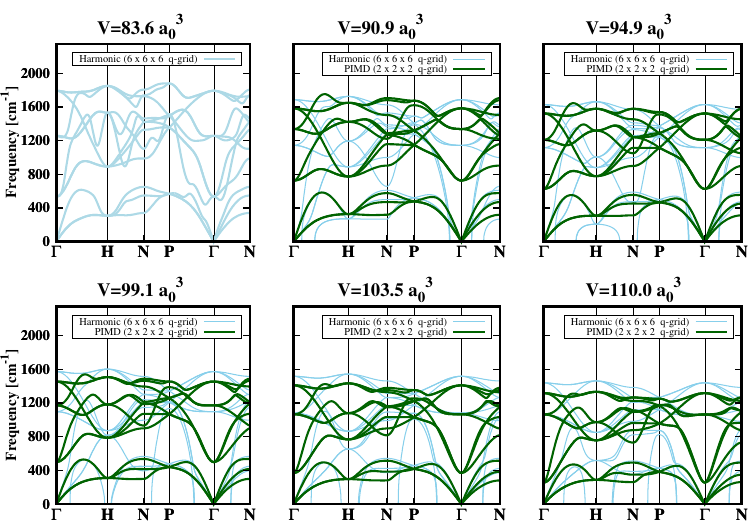}
\caption{Phonon dispersion for different volumes of the cubic Im$\bar{3}$m unit cell, whose electronic structure has been computed with the BLYP functional. At V=83.6 a$^3_0$ (volume per H$_3$S unit), only the harmonic dispersion is reported. For all the other volumes, we compare the harmonic dispersion (light-blue color) with the PIMD one (green color). Full dispersions are obtained by interpolating harmonic (anharmonic) dynamical matrices defined on a 6x6x6 (2x2x2) \textbf{q}-grid.
PIMD simulations are performed at $T$=200 K.
}\label{fig:phonons}
\end{figure*}

In Fig.~\ref{fig:phonons}, we report the analysis of the phonon dispersion for different volumes of the cubic Im$\bar{3}$m unit cell, obtained at the \emph{ab initio} level using the BLYP functional, either through the harmonic approximation via Density Functional Perturbation Theory (DFPT), or with the inclusion of quantum anharmonicity via PIMD simulations. Hereafter, volumes and energies will be expressed per H$_3$S unit, while the unit cell will be taken as cubic with S atoms 
arranged
in a bcc lattice.

At this point, it is important to underline that the DFPT and the PIMD phonons 
bear different information (see also Sec.~\ref{Sec:Phonons}).
The PIMD phonons are computed through the quantum displacement-displacement correlator recently developed in Ref.~\cite{Morresi_2021}. They describe the lowest vibrational excitations \cite{Morresi2022}, that is the energy difference between the first excited state and the ground state of the nuclear Hamiltonian.
This is the quantity normally measured by experimental probes, such as infrared or Raman spectroscopies.
Consequently, 
phonons computed in this way fully include anharmonic effects and are always positive definite, meaning that they cannot describe dynamical instabilities via the appearance of imaginary phonons.
This is at variance with
the harmonic case or with
approximated theories devised to deal with NQE, 
such as the SSCHA \cite{Monacelli2021}, which instead provide information about the sign of the free energy curvature at the reference geometry.

\begin{figure}[htb!]
    \centering
    \includegraphics[width=0.9\linewidth]{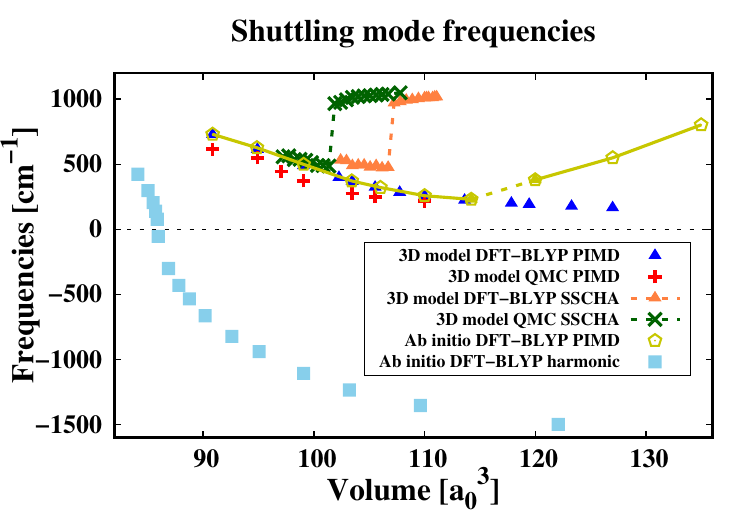}
    \caption{Shuttling mode frequencies as a function of volume by different approaches. At the PIMD level, the phonon frequencies are computed using the S-S midpoint
    as reference position for the quantum displacement-displacement correlator \cite{Morresi_2021}. Within the SSCHA framework, phonons are computed using the centroid position obtained through the free energy minimization and the full self-energy dynamical correction is included \cite{Monacelli2021}. Imaginary phonon components are represented with negative values.}\label{fig:transition_shuttle_phon}
\end{figure}

While for V=83.6 a$^3_0$ 
only 
the harmonic dispersion 
is reported in Fig.~\ref{fig:phonons}, 
for 
larger volumes we compare the PIMD phonons (green lines) obtained in a 2x2x2 supercell with the harmonic ones (light-blue lines). 
We notice that for PIMD phonons, the spatial range of the force constant matrix is such that the 2x2x2 supercell is large enough to allow for a $\mathbf{q}$-interpolation of the phonon branches $\omega_m=\omega_m(\mathbf{q})$. The comparison between PIMD and harmonic phonons of Fig.~\ref{fig:phonons} clearly shows how strong NQE are and how sizable is the softening of the most energetic phonons due to quantum anharmonicity, particularly at the largest volumes.
In the harmonic framework, for V$>$85 a$_0^3$ (see Figs.~\ref{fig:phonons} and \ref{fig:transition_shuttle_phon}), 
the appearance of imaginary frequencies indicates the dynamical instability of the Im$\bar{3}$m 
structure.
More specifically, the softening of the shuttling hydrogen mode at $\mathbf{q}=\Gamma$ signals the transition towards the 
asymmetric R3m phase \cite{Errea_2016}.
From Fig.~\ref{fig:phonons}, one can see that
imaginary frequencies disappear in PIMD phonons and their evolution 
as a function of volume is much smoother than in the harmonic case.  
As expected
from the definition of PIMD phonons,
PIMD simulations
never yield 
imaginary frequencies for the shuttling mode. 
In this regard, see also Fig.~\ref{fig:transition_shuttle_phon}, where we report the shuttling mode frequency we obtained
as a function of volume at different levels of theory. This analysis
\change{shows}
that 
\change{the putative transition implied by the maximum in T$_c$}
cannot be 
determined
using solely the shuttling mode frequency as a proxy,
\change{because in the volume range corresponding to the experimental
  T$_c$ maximum\cite{Drozdov_2015},
  i.e. between 98 $a_0^3$ and 100 $a_0^3$,\footnote{See the equations of
    state reported in Fig.~\ref{fig:drozdov_mod}(a) for the volume estimate,
    corresponding to the experimental pressure of $\approx$ 150 GPa.} there is no
  anomalous behavior of the proton shuttling mode frequency.}
We need to rely upon other observables in a framework describing nuclei as quantum particles.

So far, we have reported the structural behavior as a function of volume, fixed in our simulations. However, we can easily deduce the corresponding pressure by deriving the 
equation of state (EOS) $P=P(V)$ using the Vinet 
relation \cite{Vinet_EOS} 
computed with the same functionals 
employed
to calculate the phonon dispersions. Nevertheless, our goal is to go beyond DFT and reach a more accurate electronic description of the system using QMC methods (the details of our QMC calculations are reported in Sec.~\ref{Sec:elstructurePES}). A simple comparison of the EOS produced by the two approaches, shown in Fig.~\ref{fig:drozdov_mod}(a), reveals visible differences, suggesting that a description of the electronic structure at the QMC level is crucial to estimate correctly the critical pressure. Unfortunately, QMC calculations are much more expensive than DFT, and 
coupling them
with \emph{ab initio} PIMD simulations to study 
the real crystalline system is out of reach. Therefore, we need a simplified 
PES
describing the hydrogen shuttling mode that can be derived, after a fitting procedure, from QMC total energy calculations performed on a coarse grid of nuclear configurations. This model PES can then be used to compute the 
shuttling mode frequencies 
and to study the
\change{local polarization properties induced by the proton displacement}
at the PIMD level. 

\subsection{Classical 3D model}

The model PES is derived by considering the collective and coherent motion of all the hydrogen atoms along the direction connecting the two 
S atoms flanking each H (S-S direction), by allowing also 
hydrogen
out-of-axis mobility, while the S atoms are pinned in their bcc
positions. In this way, we aim at reproducing the shuttling mode
dynamics that takes place at $\mathbf{q}=\Gamma$, thus having the same
modulation for all hydrogen atoms in the crystal. Therefore, we reduce
the 3$N$ dimensions of the \emph{ab initio} potential (with $N$ the
number of atoms in the supercell) to only 3 dimensions.
The PES is 
fitted over total energies generated either by DFT-BLYP or by QMC for nuclear configurations defined on a cylindrical grid. Further details about the model description can be found in Sec.~\ref{Sec:Pes}.

\begin{figure*}[t!]
    \centering
    \text{\rotatebox{90}{\tiny \textbf{\hspace{0.3cm}DFT-BLYP}}}
    \includegraphics[width=0.98\linewidth]{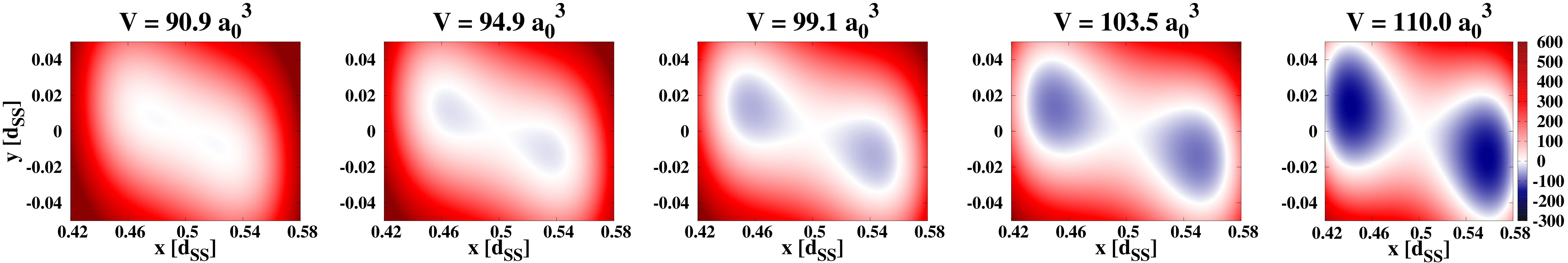}
    \text{\rotatebox{90}{\tiny \textbf{\hspace{0.8cm}1D}}}
    \includegraphics[width=0.98\linewidth]{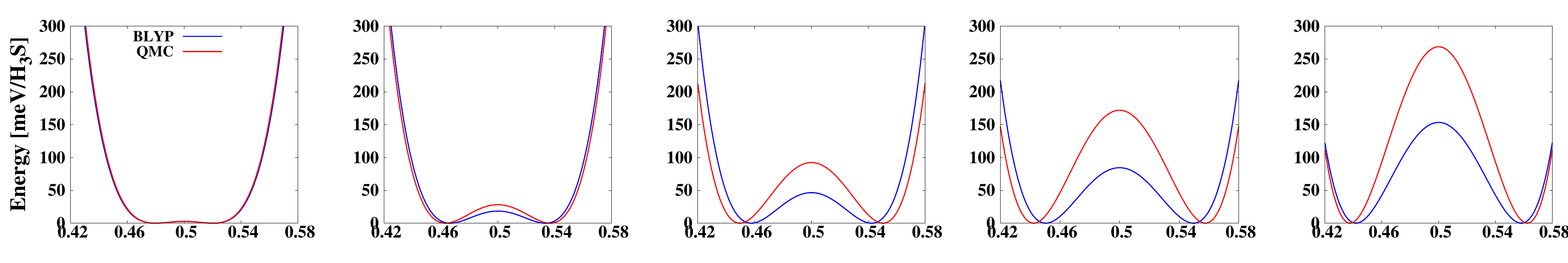}\\
    \text{\tiny \rotatebox{90}{\textbf{\hspace{0.6cm} QMC}}}    \includegraphics[width=0.97\linewidth]{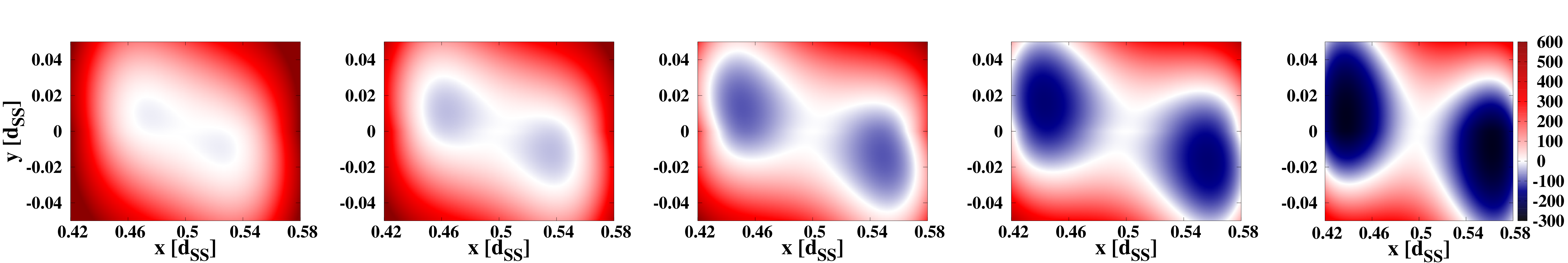}
    \caption{ 3D-PES 
    landscapes (in meV/$\ce{H_3S}$ units) for different volumes, computed on the plane containing the S-S direction ($x$ axis) and bisecting the $y$-$z$ quadrant ($y=z$ plane).
    The points on the plane are fully defined by their $x$ and $y$ coordinates, expressed in fractional units.
    Top panel: DFT-BLYP PES. Middle panel: 1D projection along the axis connecting the two minima. Bottom panel: QMC PES. In the top and bottom panels the zero of energy is the PES value at (0.5,0,0). In the middle panel, the zero is the PES minimum.}
    \label{fig:pes_qmc_dft}
\end{figure*}

In Fig.~\ref{fig:pes_qmc_dft}, we report the PES profiles obtained by
solving the electronic problem
within the DFT-BLYP (first row) and QMC (third row) methods.
At the volumes taken into account here, both DFT-BLYP PES and QMC PES have two minima connected through the inversion symmetry with respect to the S-S midpoint ((0.5,0,0) in fractional units).
The second row shows a comparison of both energy profiles cut along the line connecting these two PES minima, 
going
through the S-S midpoint. 
For the smallest volume analyzed, V=90.9 a$_0^3$, we found a good agreement between the DFT-BLYP PES and QMC PES, suggesting that electron correlation effects are reasonably well described at the DFT level at high-enough pressures. 
However,
the discrepancy between the two approaches 
appears
when we increase the volume and it grows continuously upon 
pressure release.
For the largest volume considered, V=110.0 a$_0^3$, the height of the double well barrier for QMC is $\sim$270 meV/$\ce{H_3S}$, 80 \% larger than the DFT-BLYP one.

The \change{ferroelectric} transition volume for classical nuclei can be estimated based on the PES by using the Landau theory for continuous phase transitions \cite{Landau_1937}. This method relies on the sign change of the free energy curvature (the total energy curvature at $T$=0 K) at the volume when the two displaced minima merge into a single one, in the symmetric configuration corresponding 
to the point (0.5,0,0) in Fig.~\ref{fig:pes_qmc_dft}.
For DFT-BLYP, we found a critical volume \change{$V_\textrm{ferro}$} around 85 $a_0^3$ corresponding to a pressure of 263 GPa, while for QMC we found the same volume ($\approx$ 85 $a_0^3$) which corresponds to 238 GPa in this case (see Fig.~\ref{fig:drozdov_mod}(a)). We note that the 
\change{$V_\textrm{ferro}$} yielded by the BLYP 3D-PES is in nice agreement with the value at which the shuttling mode frequency vanishes, computed 
\emph{ab initio} 
in the harmonic approximation (see Fig.~\ref{fig:transition_shuttle_phon}). This is a signature of our model PES quality.

\subsection{Quantum 3D model}

In order to have a reliable description of the structural phase transition based on our 3D-PES, we need
to include nuclear quantum effects. We add them by performing PIMD calculations
as implemented in Ref.~\cite{Mouhat_2017}. Numerical details of these
simulations can be found in Sec.~\ref{Sec:PIMD}.
\change{Here, we mention only that in the PIMD simulations of our 3D model the hydrogen atom has an
 effective mass equal to three times the physical hydrogen mass, owing
 to the fact that the PES is expressed per H$_3$S unit and the 3D motion of all hydrogen atoms in the H$_3$S
molecule is concerted by construction.}

\begin{figure*}[t!]
\includegraphics[width=1.\textwidth]{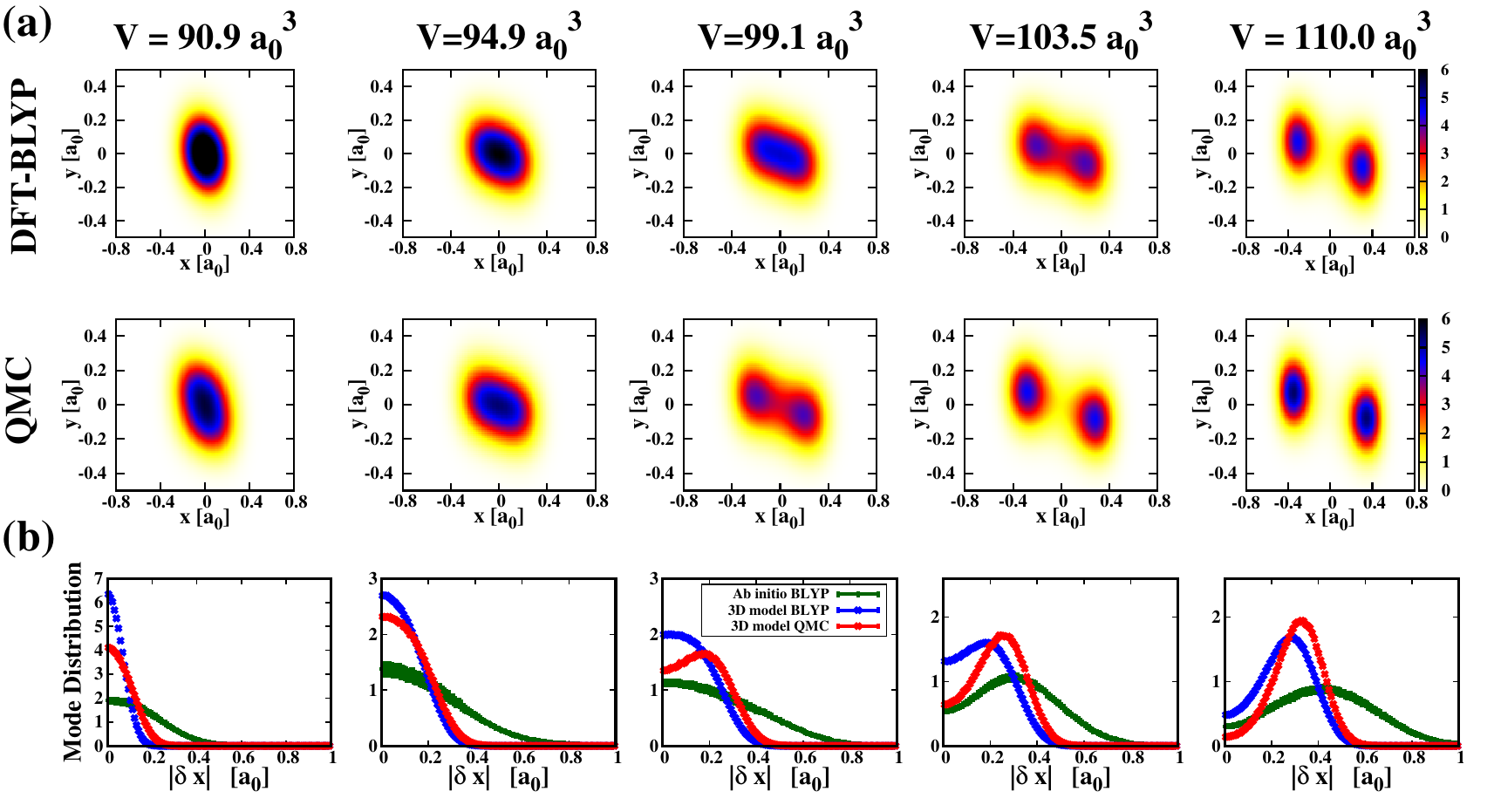}
    \caption{(a) Two-dimensional projection on the $y=z$ plane of hydrogen positions with our 3D model solved by PIMD. The results of BLYP PES calculations are represented in the top panel and the QMC PES ones in the bottom. (b) Hydrogen distribution projected along the shuttling mode. Blue points are for the BLYP 3D-PES; green points for the \emph{ab initio} BLYP simulations; red points for QMC 3D-PES. Lengths are expressed in Bohr. The origin of the reference frame is centered at the S-S midpoint. $\lvert \delta \mathbf{x} \rvert$ is the distance from the origin.}
    \label{fig:distributions}
\end{figure*}

In Fig.~\ref{fig:distributions}, we report the projections of the resulting 3D proton density, 
which takes two distinct 
shapes
depending on the volume.
The density
exhibits only one peak centered in the middle of the S-S axis
\change{for small volumes} and
the central peak \change{splits} into two lobes for the largest volumes. 
In the contour plot of Fig.~\ref{fig:distributions}(a), one can clearly see that the doubling of the peak happens in QMC at smaller volumes (higher pressures) than in DFT-BLYP, as expected from the analysis of the classical PES, which shows deeper minima in the QMC PES at fixed volume. In Fig.~\ref{fig:distributions}(b), we plot the distribution of the hydrogen position projected along the shuttling mode direction, by including also the data coming from the \emph{ab initio} PIMD simulations driven by DFT-BLYP forces. Our 3D model has a
similar behavior in comparison with the 
full 
3$N$ dimensional system.
The mode distribution assumes a double peak shape at approximately the same volume for the model (blue lines) and the \emph{ab initio} system (green lines), evaluated for the same BLYP functional. The main differences are the broadness of the distribution, underestimated by the model, and the position of the peak, which lies closer to the S atoms in the 
\emph{ab initio} simulations. 
These differences can
be understood based on the enhanced quantum-thermal fluctuations of
the \emph{ab initio} system compared to the one with a reduced number
of degrees of freedom. Nevertheless, as far as the
\change{the peak splitting} is concerned, the \emph{ab initio} and the model PIMD calculations are in agreement. This 
validates the accuracy of our 3D-PES model, which then allows one to compare directly 
BLYP and QMC results. The projected 1D distribution in
Fig.~\ref{fig:distributions}(b) reveals that the QMC PES leads to a
smaller
volume \change{for the peak splitting}, as shown already in the contour plot of Fig.~\ref{fig:distributions}(a).

By fitting the 
distribution in Fig.~\ref{fig:distributions}(b) and interpolating the parameters obtained for several volumes, it is possible to determine precisely the position of its maximum as a function of volume, and thus the occurrence of 
\change{the bimodal distribution}. 
Moreover, in PIMD we can easily quantify isotope effects
by replacing the hydrogen with the deuterium mass.

We estimate the
\change{peak splitting} to take place in $\ce{H_3S}$ at a volume of
99.6 $a_0^3$ for DFT-BLYP, and of 96.3 $a_0^3$ for QMC.
According to the EOS of
Fig.~\ref{fig:drozdov_mod}(a), these volumes correspond to
pressures of 153 GPa for DFT-BLYP and of 152 GPa for
QMC. These values, reported in
Tab.~\ref{tab:comparison_fluc_dens_d3s_h3s_qmc_blyp}, are in good
agreement with the position of the maximum $T_c$ measured in
experiments \cite{Drozdov_2015}, \change{and they are strongly
  affected by NQE.}
Nevertheless, it is important to underline that the two similar
pressures obtained by BLYP PES and QMC PES after
inclusion of NQE originate from a compensation of errors in the BLYP
values, if we take QMC as reference. Indeed, a
volume
overestimation found in the BLYP PES compensates with a pressure
overestimation in the BLYP EOS to yield approximately the same
pressure \change{for the peak splitting} as the one found in QMC (see Fig.~\ref{fig:drozdov_mod}(a)).

\begin{figure*}[hbt!]
    \centering
    \includegraphics[width=1.0\textwidth]{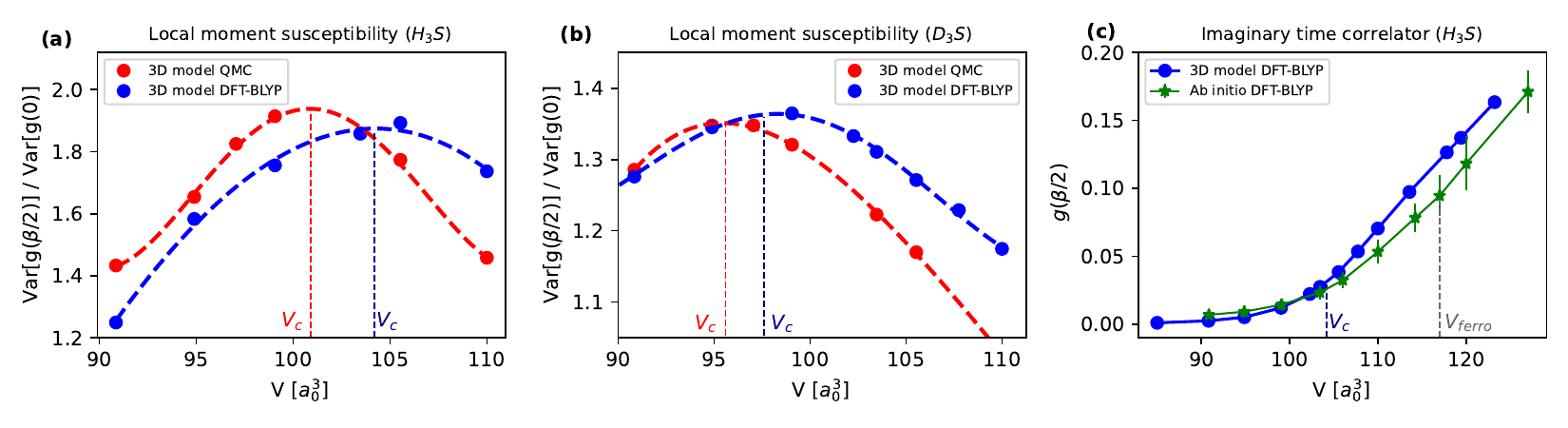} 
    \caption{\change{Local moment susceptibility for $\ce{H_3S}$ (panel (a)) and $\ce{D_3S}$ (panel (b)) for quantum nuclei with the DFT-BLYP (blue) and QMC (red) 3D-PES. The vertical dashed lines indicates the susceptibility maxima. Panel (c): Imaginary time correlator $g(\beta/2)$ for $\ce{H_3S}$. Blue circles and green stars are DFT-BLYP results for the 3D-PES and \emph{ab initio} simulations, respectively. For reference, vertical dashed lines indicate the position of the local moment susceptibility peak and the ferroelectric transition, as determined in Fig.~\ref{fig:chi_abinitio}.}}
    \label{fig:susceptibility}
\end{figure*}

\change{The occurrance of the bimodal distribution signals the
  proximity of a critical region where the quantum proton is  more localized, either
  dynamically
  or statically, in one of the two
  wells. However, from a more rigourous point of view the
  instantaneous localization of the proton, namely the local
  dipole
  formation, is better characterized by}
the ``local moment'' susceptibility, \change{as defined
  below}. Indeed, quantum fluctuations are at work across the
transition to make the hydrogen shuttle between the two PES minima. As
the volume increases and the minima deepen, the fluctuations will
start freezing, leading to the creation of a local
\change{electric dipole} moment,
generated by the \change{statically} displaced proton in the R3m phase, \change{or generated dynamically, by instantaneous configurations where the whole path representing the quantum proton is fully localized in one of the two wells.} PIMD fully accounts for quantum fluctuations, thanks to its imaginary time resolution. We can measure them by computing the imaginary time correlator $ g(\beta/2) = \langle\delta x (0) \delta x(\beta/2)\rangle$, with $\beta = 1/(k_B T)$ the inverse temperature used in the PIMD simulations, and $\delta x (\tau) = x(\tau) - \left\langle x \right\rangle$, where $\left\langle x \right\rangle$ is the thermal quantum average of the $x$ coordinate\change{, corresponding to the symmetric position in the 3D model}. Quantum fluctuations reduce the value of $g(\beta/2)$. A non-zero value of $g(\beta/2)$ can be interpreted by the presence of a finite moment in the 
distribution. In our 3D-PES, this moment is by definition local,
because by model construction the hydrogen dynamics is condensed in a
single 3D site. Therefore, the local moment susceptibility $\chi_g$ is
the normalized variance of $g(\beta/2)$, namely
$\chi_g=\textrm{Var}[g(\beta/2)] / \textrm{Var}[g(0)]$. Within the
local moment fluctuation picture, the
\change{occurrence of local polarization}
can then be estimated by evaluating the volume at which $\chi_g$ is
maximum, as shown in
\change{Fig.~\ref{fig:susceptibility}(a-b)}. This quantity has already been
used in a previous work \cite{Miha} to identify the transition from a
\change{paraelectric}
phase to a disordered regime in an anharmonic oscillator chain,
characterized by a tunable double well potential, where the symmetry
is locally \change{and instantaneously} broken in favor of displaced configurations.
If we describe the
\change{change of regime} based on local moment fluctuations,  we obtain $V_c$ = 104.2 $a_0^3$ for DFT-BLYP, corresponding to $P_c$ = 131 GPa, and $V_c$ = 100.9 $a_0^3$ for QMC, corresponding to $P_c$= 126 GPa (Tab.~\ref{tab:comparison_fluc_dens_d3s_h3s_qmc_blyp}). Also in this case, like for the density probe, cancellation of errors is at play and, by consequence, the two electronic descriptions provide 
almost
the same critical pressure. 

\change{We evaluated $g(\beta/2)$ also in our \emph{ab initio} PIMD simulations, and in Fig.~\ref{fig:susceptibility}(c) we compare it against the values of $g(\beta/2)$ coming from the PIMD solution of the 3D model. The \emph{ab initio} and model results are in statistical agreement for this local quantity, confirming that the 3D model correctly captures the volume evolution of the local polarization.}

\begin{table}[b!]
\centering
\begin{tabular}{|c|cccc|cccc|}
\hline
\textbf{Theory}     & \multicolumn{4}{c|}{\textbf{DFT-BLYP}}                                                       & \multicolumn{4}{c|}{\textbf{QMC}}                                                            \\ \hline
\textbf{Isotope}    & \multicolumn{2}{c|}{$\ce{H_3S}$}                             & \multicolumn{2}{c|}{$\ce{D_3S}$}        & \multicolumn{2}{c|}{$\ce{H_3S}$}                             & \multicolumn{2}{c|}{$\ce{D_3S}$}        \\ \hline
\textbf{Approach}   & \multicolumn{1}{c|}{Fluc.} & \multicolumn{1}{c|}{Dens.} & \multicolumn{1}{c|}{Fluc.} & Dens. & \multicolumn{1}{c|}{Fluc.} & \multicolumn{1}{c|}{Dens.} & \multicolumn{1}{c|}{Fluc.} & Dens. \\ \hline
$V_c$ {[}$a_0^3${]} & \multicolumn{1}{c|}{104.2} & \multicolumn{1}{c|}{99.6}  & \multicolumn{1}{c|}{97.6}  & 96.6  & \multicolumn{1}{c|}{100.9} & \multicolumn{1}{c|}{96.3}  & \multicolumn{1}{c|}{95.6}  & 93.6  \\ \hline
$P_c$ {[}GPa{]}     & \multicolumn{1}{c|}{130}   & \multicolumn{1}{c|}{153}   & \multicolumn{1}{c|}{165}   & 171   & \multicolumn{1}{c|}{126}   & \multicolumn{1}{c|}{152}   & \multicolumn{1}{c|}{156}   & 169   \\ \hline
\end{tabular}
\caption{\change{Transition} pressures and volumes for
  the \change{local moment formation} yielded by PIMD according to the electronic description (DFT-BLYP or QMC), the probe used (density or local moment fluctuations), and the isotope considered ($\ce{H_3S}$ or $\ce{D_3S}$).}
\label{tab:comparison_fluc_dens_d3s_h3s_qmc_blyp}
\end{table}

The two probes we used in this work, \change{the local moment
  susceptibility and the peak splitting}, allow us to determine a lower and
an upper bound for the pressure where the fluctuating local
  dipoles disappear
  in favour of a paraelectric phase, by
  squeezing the compound. Notice that this does not correspond to the ferroelectric
  transition pressure, associated instead with the global
  Im$\bar{3}$m-R3m symmetry breaking, and long-range dipole order,
  which happens at lower values.
The same analysis is carried out for both the $\ce{H_3S}$ and $\ce{D_3S}$ crystals, to estimate the magnitude of isotope effects. We summarize the results in Tab.~\ref{tab:comparison_fluc_dens_d3s_h3s_qmc_blyp}, 
where we show that the hydrogen-to-deuterium substitution brings about
an increase of the
\change{local polarization formation} pressure that falls into the [17-35] GPa range.

\change{\subsection{Full BLYP-PIMD solution of the H$_3$S phase diagram at 200
    K and comparison with SSCHA}}

\change{
  After having analyzed the local moment formation with the help of
  the 3D model, we turn now the
  attention to the ferroelectric transition, associated with the
  global R3m
  $\rightarrow$ Im$\bar 3$m transformation.
  The suitable order parameter to identify this transition is  $\Delta
  =  \langle \frac{1}{N} \sum_{i=1}^N \delta x_i \rangle$, where the sum runs over all
  the $N$ hydrogen atoms in the supercell, and $\delta x_i$ is the
  distance of the $i$-th proton from the S-S midpoint at a given
  snapshot. An equivalent
  order parameter, showing usually less statistical fluctuations, is
  $\Delta_\text{abs} =  \langle \lvert \frac{1}{N} \sum_{i=1}^N \delta
  x_i \rvert \rangle$. The brackets indicate
  the average over the classical or quantum nuclear distribution. In
 PIMD simulations an additional average is then
 done over the beads positions. From these definitions, it is clear
 that this order parameter can only be computed in our \emph{ab
   initio} simulations, being the 3D model local.
 }

\begin{figure}[t!]
    \centering
    \includegraphics[width=1.0\linewidth]{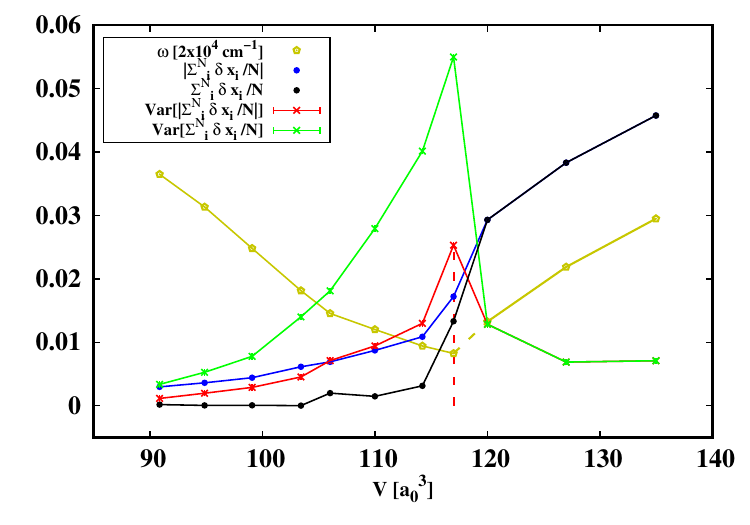}
    \caption{\change{Ferroelectric transition from \emph{ab initio}
        BLYP PIMD simulations at 200K in a 2$\times$2$\times$2
        supercell.  Black (blue) points are values of the 
        ferroelectric order parameter $\Delta$ ($\Delta_\textrm{abs}$)
        as a function of volume. The variance (i.e. susceptibility) of these order
        parameters is represented by green and red crosses,
        respectively.  We also report the shuttling mode frequencies (gold pentagons).
      }
    } 
    \label{fig:chi_abinitio}
\end{figure}

\begin{figure}[t!]
    \centering
    \includegraphics[width=1.0\linewidth]{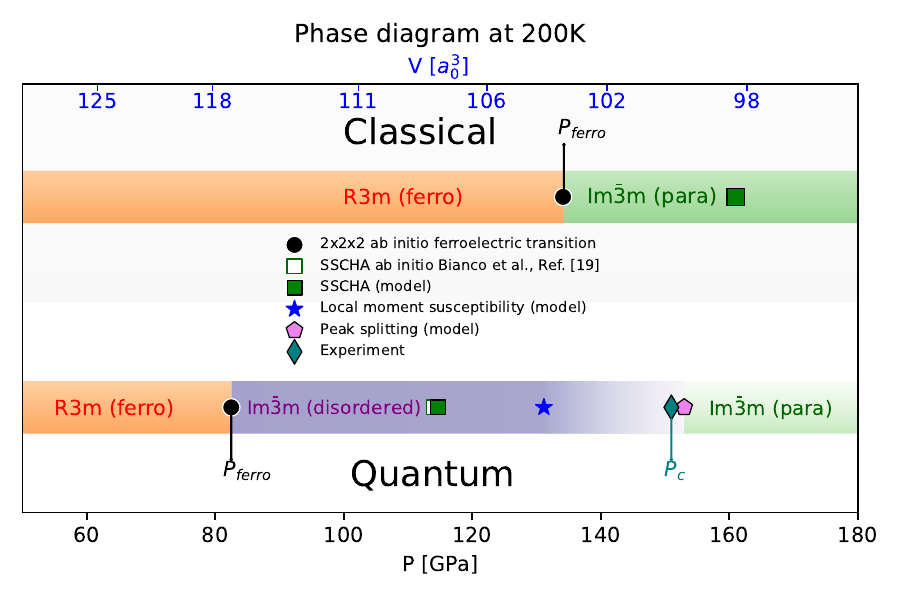}
    \caption{\change{Phase diagram at 200K for classical and quantum
        nuclei as determined by BLYP-driven classical and path
        integral molecular dynamics. Both ferroelectric transition
        (black dots) and
        the region where the local polarization vanishes into a
        paraelectric $Im\bar{3}m$ phase (bracketed by the blue star
        and the purple pentagon) are reported. Between ferro and para, an
        $Im\bar{3}m$ phase with disordered local moments is expected
        to take place\cite{Miha}. We also report the ferroelectric
        critical pressures predicted by SSCHA, from both
        \emph{ab initio}\cite{Bianco_2018} (white square) and our
        3D-PES model (green squares). For comparison, the
        pressure of the experimental $T_c$ maximum is shown by a teal diamond.        
      }  }
    \label{fig:phasediagram}
\end{figure}

\change{ 
In Fig.~\ref{fig:chi_abinitio}, we plot the volume dependence of the
order parameter $\Delta$ and $\Delta_\text{abs}$ and their
susceptibilities. Their peak is located at the
ferroelectric transition, occurring at $V_\textrm{ferro} = 117 a_0^3$,
$P_\textrm{ferro} \simeq$ 82 GPa, as found in our BLYP-PIMD simulations in the
2$\times$2$\times$2 supercell.
We notice that the peak location is correlated
with the jump in the shuttling mode frequency, reported in both
Fig.~\ref{fig:transition_shuttle_phon} and \ref{fig:chi_abinitio}.
The agreement between the  $\Delta$ and $\Delta_\text{abs}$
susceptibilities and the shuttling mode frequency jump strengthens the
reliability of our estimate. 
Interestingly, at 200K the ferroelectric transition takes place at a pressure
much lower than the one where the local moments are suppressed.
For quantum nuclei,
between these two pressures the
system is in a
regime characterized by disordered local
moments and Im$\bar{3}$m symmetry (see Fig.~\ref{fig:phasediagram} for
the resulting phase diagram).
Accounting for thermal and quantum effects leads to a strong reduction
of the critical ferroelectric pressure observed in the classical
framework, which is as large as 263 GPa at zero temperature.
Furthermore, in order to distinguish between anharmonicity coming from
thermal and NQE, we also performed classical \emph{ab initio} MD
simulations at 200K (see Supplementary Note V of the Supplementary Information (SI)),
which yield a ferroelectric transition at $\approx$
133 GPa (see Fig.~\ref{fig:phasediagram}). Thus, classical anharmonicity accounts for
about 70$\%$ of the total pressure reduction of the ferroelectric
transition at 200K. The remaining 30 $\%$ is due to NQE at the same temperature.
}

\change{
  The ferroelectric transition
cannot be
estimated
at the
QMC level, due to its computational cost when applied to the dynamics of a real
extended system. However, as we have seen, the 3D model,
derived at both the DFT-BLYP and QMC levels, is
enough to determine the formation of local electric dipole moments,
whose pressure $P_c$ matches well the position of the experimental $T_c$ maximum.
}
At variance with previous state-of-the-art calculations based on a
combination of DFT-BLYP and SSCHA frameworks, our PIMD results yield
$P_c$ in a substantial agreement with the experimental finding, and
this irrespective of the electronic theory used to generate the PES.
\change{One should notice here  that the original SSCHA approximation is not able to
  capture the disordered $Im\bar{3}m$ phase, being a mean-field theory
  with no configurational entropy and no direct information of imaginary time correlations, key to detect
  the 
  dynamical
  local moment formation\footnote{Note, however, that time resolved
    extensions of SSCHA have been recently proposed, able in principle to access also retardation effects\cite{Monacelli_2021_PRB}.}. Thus, the critical pressure SSCHA can normally compute
  is the ferroelectric one, $P_\textrm{ferro}$, and not $P_c$.
  }
  To investigate more deeply
  \change{this mismatch,}
  we carry out SSCHA calculations with our model PES (see
  Sec.~\ref{Sec:SSCHA} for details).
  In SSCHA, the occurrence of the asymmetric R3m phase is signalled by
  a centroid displaced with respect to the S-S midpoint.
  The SSCHA 
critical values are $V_\textrm{ferro} = 107.8$ $a_0^3$, $P_\textrm{ferro} = 114$ GPa for the DFT-BLYP PES, and $V_\textrm{ferro} = 102.4$ $a_0^3$, $P_\textrm{ferro} = 118$ GPa for the QMC PES. As in PIMD, there is no significant difference between the electronic structure method used to generate the PES.
Our SSCHA results for 
the DFT-BLYP PES
are in a very good agreement with the outcome of previous SSCHA simulations for the full \emph{ab initio} system \cite{Bianco_2018}, calculated with the same DFT-BLYP functional. 
We find that the SSCHA
\change{overestimates} $P_\textrm{ferro}$ with respect to the one
obtained in PIMD for the same
\change{BLYP functional}, \change{as expected from a
  mean-field theory, and underestimates $P_c$, which is however out
of reach by the SSCHA,} suggesting that the approximated description
of the nuclear Hamiltonian
\change{is the source of}
disagreement with both PIMD and experimental results.

Let us look 
now
at the predictions for the shuttling mode frequencies, plotted in
Fig.~\ref{fig:transition_shuttle_phon} for various methods. It has to
be noted that, within the SSCHA, the phonon frequency of the shuttling
mode shows a jump at \change{$V_\textrm{ferro}$}.
This is due to the 
hop of
the SSCHA centroid 
from
its symmetric position
to a different minimum of the free energy,
already
``preformed'', 
which breaks the symmetry and 
becomes energetically more favorable at $V_\textrm{ferro}$.
Moreover, we also observe an increase of the SSCHA phonon line-width across the transition of the order of 10 cm$^{-1}$. 
\change{A similar jump in the shuttling phonon frequencies is
  detected by our \emph{ab initio} BLYP simulations in correspondence
  with the ferroelectric transition (see Figs.~\ref{fig:transition_shuttle_phon} and \ref{fig:chi_abinitio}).
  }
\change{Nevertheless,} our PIMD phonon determination, 
shows
a progressive phonon softening
without jumps across the
\change{volume region of local
  moment formation}. 
This is not only true within our 3D model PES, but also for our PIMD
calculations driven by \emph{ab initio} forces computed at the
DFT-BLYP level, as shown in
Fig.~\ref{fig:transition_shuttle_phon}. The agreement between
shuttling mode frequencies yielded by the 3D model and the ones given
by \emph{ab initio} calculations \change{in this volume region} highlights once again the quality of our model PES.
\change{This supports the hypothesis of two different
  transitions. The first one is a smooth transition, or crossover, from the paraelectric
  Im$\bar{3}$m to a phase sharing the same  Im$\bar{3}$m
  symmetry and characterized by the formation
  of local and spatially disordered local moments. This phase 
  cannot be detected by looking at the phonon
  frequencies, and it is not accessible within
  the SSCHA formulation. The second one is the ferroelectric transition from the
  disordered Im$\bar{3}$m to the asymmetric R3m phase, 
  which happens at significantly lower pressure than the
  first one, where the shuttling phonon frequency shows a jump.
  The phase diagram deducible from our combined \emph{ab initio} and
  3D model results is drawn in Fig.~\ref{fig:phasediagram}. 
}

\section{Conclusions}
\label{sec12}

\begin{figure*}[tbh!]
    \centering 
    {\includegraphics[width=0.8\textwidth]{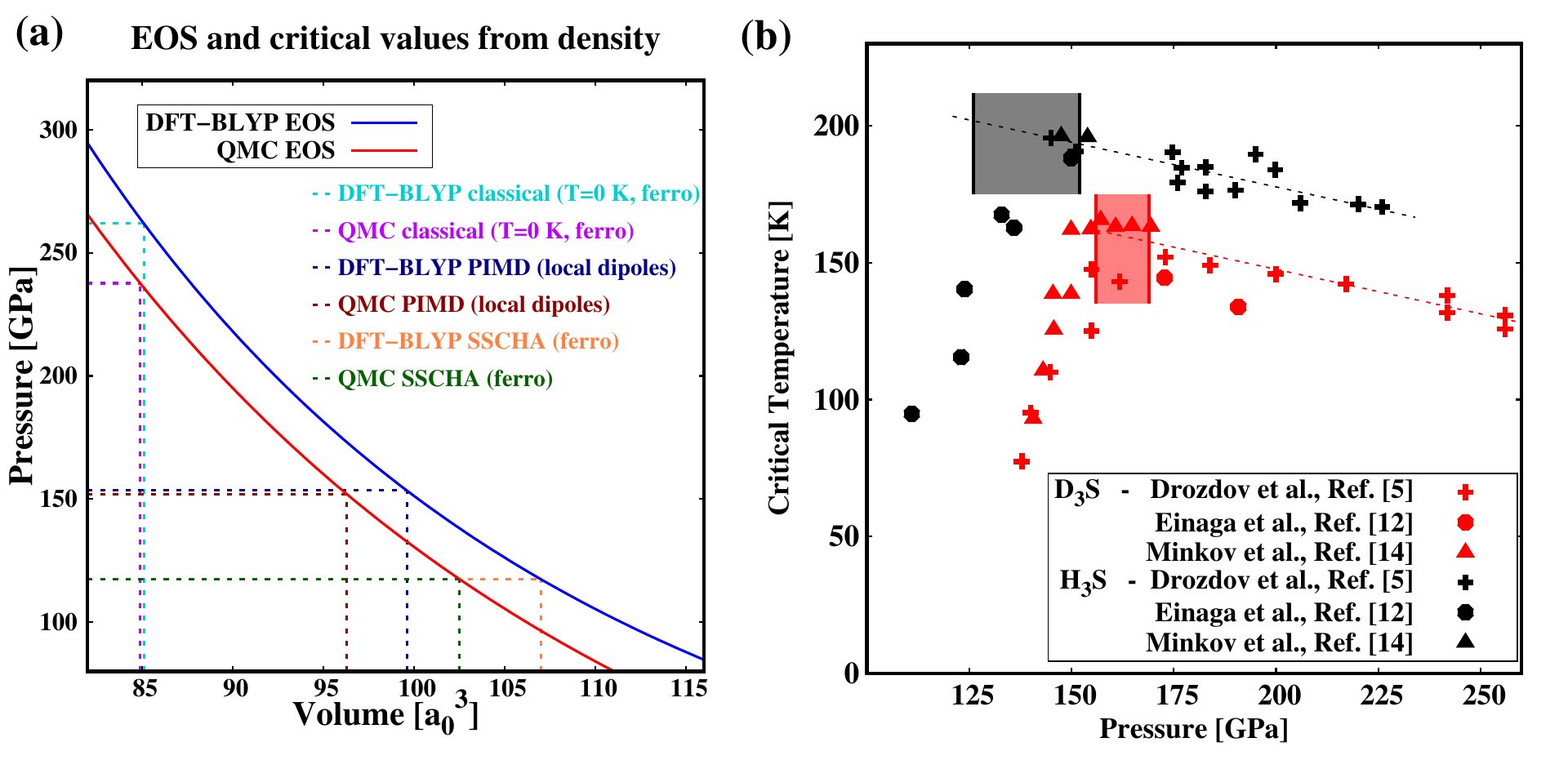}}
    \caption{(a) Equation of state $P=P(V)$ (see Sec.~\ref{Sec:EOS} and Eq.~\ref{eq:vinet_P}) for both DFT-BLYP (blue color) and QMC (red color) 
    calculations of $\ce{H_3S}$ in the Im$\bar 3$m phase.
    The transition volumes and the corresponding pressures identified at different levels of theory using the density probe are displayed by dashed lines. (b) $T_c$ 
    as a function of pressure in $\ce{H_3S}$ and $\ce{D_3S}$ from Drozdov \emph{et al.} in \cite{Drozdov_2015}, Einaga \emph{et al.} in \cite{Einaga_2016} and from Minkov \emph{et al.} in \cite{Minkov_2020}.
    The shaded areas on each data set represent the pressure range when the transition occurs according to 
    our PIMD results for the QMC PES,
    where the lower limit is 
    based on the
    local quantum fluctuations analysis and the upper one 
    on
    the density 
    evolution.
    }
    \label{fig:drozdov_mod}
\end{figure*}

In this work, starting from \emph{ab initio} electronic structure calculations, we generated a model PES to describe the shuttling mode of hydrogen in $\ce{H_3S}$, responsible for the R3m $\rightarrow$ Im$\bar 3$m transition, which was originally associated with the 
$T_c$ maximum
as a function of pressure. Despite the fact that such a hydrogen symmetrization 
is expected to happen in $\ce{H_3S}$ upon compression, so far no theoretical method has been able to spot it 
at pressures near the one that maximizes $T_c$ in experiments. This raised doubts on the original association between superconductivity and structural transition \cite{Akashi_2016,Azadi_2017}, worsened by the fact that other competing symmetries could be stable in the same pressure range \cite{Goncharov_2016,Li_2016,Guigue_2017,cui2019favored}.
The mismatch found between previous theoretical estimates of the
critical pressure $P_c$
and
the experimental values for the $T_c$ maximum is solved 
by applying
state-of-the-art computational methods in both the electronic and nuclear Hamiltonians, namely using QMC calculations for electrons, and the PIMD approach for nuclei. 
Within our QMC+PIMD approach, the experimental pressure where $T_c$ is maximum is bracketed by the $P_c$ value estimated from the local fluctuations probe and the one determined by the 
transformation
of the bimodal hydrogen distribution into a 
unimodal one.
Consequently, these two probes provide a lower and an upper bound for the critical pressure, with a range between the two of $\approx$ 20 GPa. 
The range of transition pressures identified is consistent with the available experimental data for the $T_c$ maximum \cite{Drozdov_2015,Einaga_2016,Minkov_2020} 
for both $\ce{H_3S}$ and $\ce{D_3S}$, as we can see in
Fig.~\ref{fig:drozdov_mod}(b).

We have thus shown that the occurrence of the $T_c$ maximum 
should be linked with the formation of the phase characterized by disordered local moments\cite{Miha}, and it
  cannot be associated with the ferroelectric
  R3m $\longrightarrow$ Im$\bar{3}$m transformation, which takes place at a lower pressure $P_\textrm{ferro}$ compared with $P_c$. According to our outcome, $T_c$ reaches its maximum when the local dipole moments melt upon compression, and protons become fully delocalized across the PES barrier.

Furthermore, we notice that the \emph{ab initio} electronic structure computed at the DFT-BLYP level predicts very good results for the critical pressure, similar to those obtained by QMC.
However, it is important to stress that the DFT-BLYP pressures are affected by error compensation, 
the overestimation of the critical volume 
being
balanced by a different EOS if compared against QMC calculations. This aspect underlines the importance of using an accurate electronic description, beyond the DFT level. The generation of our model PES, built to describe the hydrogen shuttling mode, allowed us to exploit the QMC energies in a PIMD framework, otherwise unfeasible in the full 3$N$ dimensional system.

We conclude by noting that the the R3m $\rightarrow$ Im$\bar 3$m structural phase transition in sulfur hydride has strong analogies with the hydrogen bond symmetrization in other compounds such as high-pressure ice, where, upon compression, 
phase VII and VIII
hosting displaced protons, 
stable at lower pressure, are expected to transform into the symmetric phase X \cite{Pruzan2003,Benoit1998}.
However, it is still a matter of debate whether the transformation is direct or whether another intermediate disordered 
structure
appears, with protons only partially symmetrized. 
In
this respect, 
further work is needed to extend our model beyond the collective path dynamics to treat non-local spatial correlations and disordered patterns. Machine learning schemes could then be useful to generate more extended PES from QMC data \cite{tirelli_2022,Huang2022,Ceperley2023} 
with the aim at including a larger variety of hydrogen configurations in PIMD calculations by keeping the same QMC accuracy.

\section{Methods}
\label{sec11}

\subsection{Electronic structure calculations for the PES model}
\label{Sec:elstructurePES}
For the DFT electronic structure calculations, we used the Quantum Espresso (QE) suite of codes \cite{qe1,qe2}, while for the QMC calculations, we employed the TurboRVB package \cite{Nakano_2020}. For sake of consistency, in both DFT and QMC calculations, we used the same set of pseudopotentials. Namely, 
we treated the sulfur atom with the ccECP neon-core pseudopotential \cite{Bennett_2017} particularly suited for correlated calculations, available in both the QE-compatible Unified Pseudopotential Format (UPF) and in the TurboRVB-compatible Gaussian expansion format. For hydrogen, we used the bare Coulomb potential, with a very short-range cutoff for a QE usage within the plane-wave framework. In the QMC calculations instead, no short-range cutoff is needed for the bare Coulomb potential, because the nuclear cusp conditions are automatically fulfilled by our QMC wave function (see below). These pseudopotentials have been chosen after performing preliminary calculations at the DFT level to test their accuracy. We also tested other pseudopotentials (ultrasoft (US), projector augmented wave (PAW), and a combination of the above), by comparing the total energy profile obtained by moving the hydrogen atom away from the S-S midpoint, and constrained to stay on
the S-S axis. This leads to a very crude one-dimensional (1D) PES, which is however useful for testing purposes, with the advantage that it is easily computable for its simplicity.
We took as reference the total DFT energy computed with the all-electron LAPW approach, as implemented in Elk \cite{elk}. The ccECP pseudopotential for the sulfur atom and the bare Coulomb potential with short-range cutoff for the hydrogen atom turned out to be the most accurate choice \change{(see Fig.~S.1 of the SI)}. 

For 
single-point calculations at selected nuclear configurations,
we carried out DFT calculations with the Becke-Lee-Yang-Parr (BLYP) functional \cite{Becke_1988,Lee_1988}. The cutoff energy for plane waves is set to 200 Ry (due to the hardness of the H Coulomb pseudopotential), with the smearing parameter equal to 0.002 Ry and a $\mathbf{k}$-points grid of 32x32x32.

For the QMC calculations, we used a Slater-Jastrow wavefunction $\Psi$, which reads as:
\begin{equation}
\label{Eq:QMC_wavefunction}
    \Psi = \Phi_{S} \cdot \exp(J),
\end{equation}
where the term $\exp\left( J \right)$ is the Jastrow factor, symmetric under electron exchange, while $\Phi_{S}$ is the antisymmetric Slater determinant. 
The Slater orbitals in $\Phi_{S}$ are generated by DFT calculations within the Local Density Approximation (LDA) 
\cite{Kohn1965}, 
performed in a Gaussian basis set by means of the DFT 
code
built in TurboRVB. For the sulfur atom, we employed a modified cc-pVTZ primitive basis set with $6s6p2d1f$ components, contracted into 11 hybrid orbitals through the Geminal Embedded Orbitals (GEO) procedure \cite{Sorella_GEO}. For hydrogen, we used a modified cc-pVTZ primitive basis set with $4s2p1d$ components contracted into 6 GEO hybrid orbitals. 

The Jastrow exponent $J$ 
introduces explicitly electronic 
correlation in the wavefunction, and it can be decomposed into three terms, such that $J  = J_1 + J_2 + J_{3}$. 

$J_1$ is the so-called one-body term, which takes into account the interaction effects between the electrons $i$ and a nucleus $I$, and it depends on the relative electron-nucleus distances $r_{iI}$. $J_2$ is the so-called two-body term, treating the correlations between
electrons $i$ and $j$, and depending on their relative distance $r_{ij}$. Both $J_1$ and $J_2$ are designed to fulfill the electron-nucleus and electron-electron cusp conditions, respectively. They read as $J_1=\sum_{i=1}^{N_e} \sum_{I=1}^{N} u_I(r_{iI})$, and $J_2=\sum_{i<j=1}^{N_e} v(r_{ij})$, where $N$ ($N_e$) is the number of nuclei (electrons) in the supercell, and the functions $u$ and $v$ are defined as follows:
\begin{eqnarray}
    u_I(r) &=& \frac{Z_I}{a} (1-e^{-ar }) \label{Eq:u_func}\\
    v(r) &=& \frac{ r }{2(1+b r)} \label{Eq:v_func},
\end{eqnarray}
with $a$ and $b$ variational parameters, and $Z_I$ the charge of the $I$-th pseudoatom.  
The coefficients in Eqs.~\ref{Eq:u_func} and \ref{Eq:v_func} are set to fulfill the Kato cusp conditions for electron-nucleus and electron-electron coalescence, respectively \cite{Kato1957}. 

$J_{3}$ is the three-body term that accounts for the electron-electron-nucleus interactions. As defined in TurboRVB, it is also intrinsically non-homogeneous, because it depends on the individual electron positions and not only on the relative distances, which is less accurate.
Being non-homogeneous, it is expanded on a modified atomic Gaussian basis set of $2s2p1d$ atomic orbitals, for both sulfur and hydrogen atoms. 

The $J_3$ parameters, together with $a$ and $b$, are optimized by minimizing the variational energy of the many-body wavefunction in Eq.~\ref{Eq:QMC_wavefunction}.
The Slater part is instead kept frozen as determined by DFT-LDA.
As stochastic minimization algorithm, we employed the linear method \cite{Umrigar_opt}.
We then carried out lattice regularized diffusion Monte Carlo (LRDMC) calculations \cite{Casula_lrdmc}, to stochastically project the initial wavefunction towards the ground state of the system, within the fixed node approximation. 
\change{Within this approximation, the LDA nodes provide accurate results for this system, as verified in Supplementary Note II of the SI.}
In LRDMC, we used a lattice space of 0.25 $a_0$, which is known to produce converged energy differences. We started the projection from the best variational state optimized in the previous step, taken as trial wavefunction. Finite-size scaling has been performed on the 2x2x1, 2x2x2, 3x2x2 and 3x3x2 real-space supercells in order to extrapolate the LRDMC total energy to the thermodynamic limit, by also using Kwee-Zhang-Krakauer (KZK) \cite{Kwee2008}  corrections to make its size dependence milder.

This workflow has been repeated for every point in the real-space grid used to interpolate the PES model from \emph{ab initio} data (see Sec.~\ref{Sec:Pes}).

\subsection{Potential energy surface parametrization}
\label{Sec:Pes}

To derive an effective low-dimensional PES, we considered the collective and concerted motion of all the hydrogen atoms of the cubic unit cell, with sulfur atoms forming a bcc sublattice. The position of a hydrogen atom is described by the cylindrical coordinates $x$, $r$ and $\phi$, defined along the axis connecting the two flanking sulfur atoms (S-S axis): $x$ is the position of the hydrogen atom along the S-S axis, $r$ is the radial distance from the S-S axis, and $\phi$ is the azimuthal angle, wrapping around the same axis. We use fractional coordinates, where the lengths are expressed in $d_\textrm{SS}$ units, $d_\textrm{SS}$ being the lattice parameter of the bcc unit cell. Within this reference system, the S-S midpoint has coordinates $(x, r, \phi) \equiv (0.5,0,0)$. We assume that all hydrogen atoms in the unit cell move in the same way. This fixes the choice of a collective path connecting the Im$\bar 3$m symmetry (with all hydrogen atoms sitting at the S-S midpoints) to the R3m one (with all hydrogen atoms coherently displaced from the midpoint). In this way, we apply a dimensionality reduction of the full potential, depending on 3$N$ dimensional coordinates, where $N$ is the number of atoms in the cell, to a much simpler 3D PES: $E=E(x,r,\phi)$.

The functional form of our 3D PES is constructed as follows:
\begin{equation}
    E(x,r,\phi) = A(x,r) + B(x,r)\sin(\phi + 5\pi/4),
    \label{Eq:E_def}
\end{equation}
with:
\begin{eqnarray}
    A(x,r) & = &
                 \frac{f_{\textrm{max}}(x,r)+f_{\textrm{min}}(x,r)}{2}, \nonumber
  \\
  B(x,r) & = & \frac{f_{\textrm{max}}(x,r)-f_{\textrm{min}}(x,r)}{2},
    \label{Eq:AB_def}
\end{eqnarray}
and where $f_{\textrm{min}}$ and $f_{\textrm{max}}$ are defined as:
\begin{equation}
\begin{split}
f_{\textrm{min},\textrm{max}}(x,r) = 
    & \ a +  \frac{1}{2}b (x-0.5)^2 + \frac{1}{4} c (x-0.5)^4\\
    &+ d r + \frac{1}{2}e  r^2\\
    &\pm f  (x-0.5) r \pm  g  (x-0.5) r^2 \\
    &\pm h_1 (x-0.5)^3  r \pm h_2 (x-0.5)^3  r^2\\
    &+ h_3  (x-0.5)^2  r + h_4  (x-0.5)^2  r^2 \\
    &+ h_5  (x-0.5)^4  r + h_6  (x-0.5)^4  r^2
\end{split}
\label{Eq:fmaxmin_def}
\end{equation}

The choice of this functional form 
is motivated by
the symmetries of the system. For a fixed $\lbrace x,r \rbrace$, the 
potential
$E$ has an angular dependence that varies following a sine curve with $2\pi$-periodicity. In particular, for $x < 0.5$, $E$ has a minimum given by $f_{\textrm{min}}$ at $\phi=\pi/4$ and the maximum $f_{\textrm{max}}$ at $\phi=5\pi/4$. This dependence is built in Eqs.~\ref{Eq:E_def} and \ref{Eq:AB_def}. The $\{f_i(x,r)\}_{i=\textrm{min},\textrm{max}}$ functions in Eq.~\ref{Eq:fmaxmin_def} are a composition of the following terms: a Landau-type potential that well describes the energy profile for $r=0$, a second-order polynomial function in $r$ for $x=0.5$, and mixed terms made of cross products of factors up to the fourth order in $(x-0.5)$ and up to the second order in $r$, which give enough flexibility in order to well reproduce the total PES. The signs in $\{f_i(x,r)\}_{i=\textrm{min},\textrm{max}}$ ensure the symmetry: $E(1-x,r,\phi+\pi)=E(x,r,\phi)$, fulfilled by the system.

We sampled the PES by discretizing the 3D space according to the following grid defined in cylindrical coordinates: $x = \left[0.42,0.44,0.46,0.48,0.5\right]$ (in $d_{SS}$ units), $r=[0.00,0.02,0.05,0.08]$ (in $d_{SS}$ units) and $\phi=[\pi/4,5\pi /4]$. For these points we computed the \emph{ab initio} total energies, given either by DFT-BLYP or by QMC calculations. We finally used the generated datasets to best fit the PES, parametrized according to Eqs.~\ref{Eq:E_def}, \ref{Eq:AB_def} and \ref{Eq:fmaxmin_def}. The root mean square error of these fits amounts to $\approx$ 1 meV/H$_3$S.

\subsection{Equation of state}
\label{Sec:EOS}

In order to get the pressure associated to each volume, $P=P(V)$, we use the Vinet EOS:
\begin{equation}
         P(V) = 3B_0 \frac{1-\eta}{\eta^2} \exp\left(-\frac{3}{2}(B_0'-1)(1-\eta)\right).
        \label{eq:vinet_P}
\end{equation}
with $\eta=(V/V_0)^{1/3}$. In Eq.~\ref{eq:vinet_P}, the parameters $V_0$, $B_0$ and $B_0'$ are the equilibrium volume, the isothermal bulk modulus, and the derivative of bulk modulus with respect to pressure, respectively. 
The Vinet EOS \cite{Vinet_EOS} is empirical and, despite having only a few parameters, it is very accurate 
to describe solids under extreme conditions.
We obtained
$V_0$, $B_0$ and $B_0'$
by fitting the $E=E(V)$ relation for the Im$\bar{3}$m phase, where the total energy is computed from first principles, either by DFT-BLYP or by QMC, on a grid of volumes (see Fig.~\ref{fig:drozdov_mod}(a)). In the fit, we disregarded the Zero-Point Energy (ZPE) contribution, because we verified that the ZPE variation is very small ($<$ 1 mHa/H$_3$S) in the range of pressures analyzed here within the same Im$\bar{3}$m phase. 

\subsection{PIMD simulations}
\label{Sec:PIMD}
The 
PIMD simulations are carried out at 200 K using 20 beads 
\change{with \emph{ab initio} DFT forces, while using 40 beads with 3D-PES forces,}
to take into account quantum effects. 
\change{A convergence study of the PIMD results with respect to the number of beads is reported in Supplementary Note III of the SI.}
Nuclei are evolved in time using the PIOUD integrator~\cite{Mouhat_2017} with a time step equal to 0.75 fs and a friction parameter of the Langevin thermostat equal to 1.46$\cdot$10$^{-3}$ atomic units. The latter value is the same as in Ref.~\cite{Mouhat_2017}, where it is found to be optimal for both stochastic and deterministic forces. Simulations lasted around 6 ps, until the convergence of the vibrational modes at $\Gamma$ is reached. Forces are computed from the Born-Oppenheimer PES evaluated at DFT level within the QE package, or from the model PES defined in Sec.~\ref{Sec:Pes}. In case of \emph{ab initio} PIMD, we used a BLYP 
functional for computing the PES. The wavefunction cut-off for the PES is set to 90 Ry (420 Ry for the charge density), while the Fermi smearing is Gaussian and set equal to 0.03 Ry. PIMD simulations are performed using 2x2x2 real-space supercells, containing in each case 32 atoms, and the corresponding reciprocal-space mesh is always equal to 9x9x9. We used a smaller plane-wave cutoff than the one used in single-point DFT calculations, because in PIMD we replaced the hard H Coulomb pseudopotential with a smoother PAW one. This has been necessary to speed up the PIMD calculations, which would otherwise have been too time consuming.

\subsection{SSCHA simulations}
\label{Sec:SSCHA}

Besides the exact description of quantum nuclear motion provided by
PIMD, one can also rely on approximated theories like the SSCHA
\cite{Monacelli2021}, based on a variational principle on the free
energy, which allows one to include quantum nuclear anharmonicity in a
non-perturbative way. Here, we performed SSCHA simulations on the 3D
$\ce{H_3S}$ ($\ce{D_3S}$) model using up to 30000 configurations. The
average proton position (centroid) reported in
\change{Supplementary Fig.~S.5 of the SI} are directly
accessible through the SSCHA free energy minimization.

\subsection{Phonons}
\label{Sec:Phonons}

Harmonic phonons are obtained through 
DFPT
simulations \cite{Baroni_2001} as implemented within QE \cite{qe2}. The same set of DFT parameters and pseudopotentials employed for PIMD simulations were used to compute harmonic phonon dispersions, except for the \textbf{k}-space grid that was chosen equal to 18x18x18, as in this case 
it is referred to
the unit cell. The results of these calculations are shown in Fig.~\ref{fig:phonons}. We specify that the DFPT shuttling mode frequency at $\mathbf{q}=\Gamma$, reported in Fig.~\ref{fig:transition_shuttle_phon} has been computed with higher precision by employing the more accurate H Coulomb pseudopotential, requiring a plane-wave cutoff of 200 Ryd.

Anharmonic phonon frequencies at PIMD level are evaluated by computing the
zero frequency component of the phonon Matsubara Green's function from PIMD simulations. This method has been recently implemented in Ref.~\cite{Morresi_2021} and it has been shown to describe accurately the vibron frequencies of solid phases of hydrogen. Conversely, within the SSCHA, auxiliary phonons are a byproduct of the free energy minimization. However, to get the physical phonons of Fig.~\ref{fig:transition_shuttle_phon}, probed by spectroscopies, we apply the full self-energy dynamical corrections to the auxiliary dynamical matrix, described in detail in Ref.~\cite{Monacelli2021}, including both the third and fourth-order terms.

\begin{acknowledgments}

The authors thank M. Calandra, I. Errea, F. Mauri and L. Monacelli for useful discussions. They acknowledge computational resources provided by GENCI under the allocation number 0906493, which granted access to the HPC resources of IDRIS and TGCC. They also thank RIKEN for providing computational resources of the supercomputer Fugaku through the HPCI System Research Project ID hp220060.
The authors are grateful to the European Centre of Excellence in Exascale Computing TREX-Targeting Real Chemical Accuracy at the Exascale, which partially supported this work.
This project has received funding from the European Unions Horizon
2020 Research and Innovation program under Grant Agreement No. 952165.

\end{acknowledgments}

\clearpage

\onecolumngrid

\setcounter{figure}{0}
\setcounter{page}{1}
\setcounter{section}{0}
\setcounter{table}{0}

\renewcommand{\thepage}{S\arabic{page}}

\renewcommand{\figurename}{{\bf Supplementary Figure}}
\renewcommand{\tablename}{{\bf Supplementary Table}}
\renewcommand\thesection{Supplementary Note \Roman{section}}
\renewcommand\thesubsection{Supplementary Note \Roman{section}.\arabic{subsection}}
\renewcommand\thesubsubsection{Supplementary Note \Roman{section}.\arabic{subsection}.\alph{subsubsection}}
\renewcommand{\theequation}{S.\arabic{equation}}
\renewcommand{\thefigure}{S.\arabic{figure}}
\renewcommand{\thetable}{S.\arabic{table}}

\section*{Supplementary information}

\section{Pseudopotentials}
\begin{figure}[h!]
    \centering
    \includegraphics[width=\linewidth]{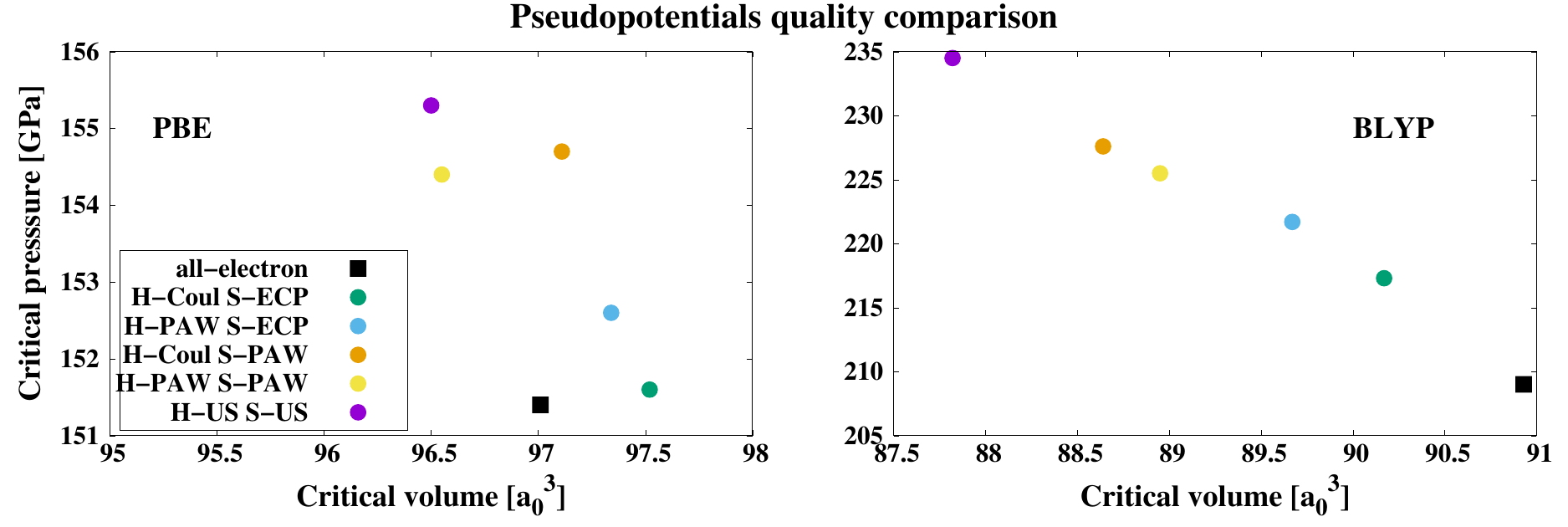}
    \caption{Pseudopotential accuracy test performed for a combination of H and S pseudopotentials, reported in the key, compared to all-electron results yielded by the Elk code\cite{elk}. The 1D model system to estimate the critical point is obtained by constraining the hydrogen atom to move on the S-S axis. We determined the classical $P_\textrm{ferro}$ (y-axis) and $V_\textrm{ferro}$ (x-axis) for the classical symmetrization transition of this 1D potential energy surface (PES), either with the PBE functional (left panel) or with the BLYP functional (right panel). 
    }
    \label{fig:pp_quality}
\end{figure}

The pseudopotentials for DFT calculations are tested against all-electron calculations, performed with the same functional. Results are shown in Fig.~\ref{fig:pp_quality} where two different functionals (PBE\cite{Perdew1996} on the left and BLYP\cite{Becke_1988,Lee_1988} on the right) have been analysed. In both cases, the bare Coulomb potential for hydrogen and the ccECP\cite{Bennett_2017} for sulfur (green points) give the best agreement with all-electron results (black squares).

\section{Impact of the fixed node approximation in diffusion Monte Carlo calculations}

As reported in the Methods section of the main paper, the Slater orbitals of the Jastrow-Slater quantum Monte Carlo wave function, defining the nodal surface, are generated by the LDA functional\cite{Kohn1965}. As we do not relax the nodal surface at the variational Monte Carlo (VMC) level, the nodes we use in the diffusion Monte Carlo (DMC) are those directly provided by the LDA Khon-Sham orbitals. 

Here, we would like to assess the quality of the LDA nodes for the H$_3$S Hamiltonian. To this aim,
we carried out selected QMC calculations, where we first relaxed the LDA nodes at the VMC level in the 2$\times$2$\times$2 cubic supercell at the Baldereschi k-point, then we performed DMC calculations with the optimized nodal surface to see the impact of the nodal surface optimization, by comparing these results  with those obtained with the frozen LDA nodes calculations. We checked the quality of the LDA nodes for both the barrier height at the volume of 110 a$_0^3$, and the equation of state (EOS) in the symmetric Im$\bar{3}$m configuration. 

As for the barrier height, we report in Tab.~\ref{tab:relaxed_det} 
the energy differences between the symmetric configuration, i.e. with the proton in the middle of the S-S segment, and a displaced one with the proton sitting at the position shifted by 0.06 $d_\textrm{SS}$ from the S-S center, quite close to the actual minimum of the double well potential at that volume.

\begin{table}[h!]
\centering
\begin{tabular}{|c|c|c|}
\hline
\textbf{Wave function} & \textbf{VMC} & \textbf{DMC} \\ 
\hline
\hline
\textbf{LDA nodes}    &  0.228(6)  &  0.261(10) \\ \hline
\textbf{relaxed determinant}  & 0.243(5) & 0.256(10) \\ \hline
\textbf{Z-relaxed determinant} & 0.245(6) & 0.262(8) \\ \hline \hline
\end{tabular}
\caption{
DMC energy differences in meV/H$_3$S between the symmetric proton configuration and the one displaced by 0.6 $d_\textrm{SS}$, with  $d_\textrm{SS} = 6.0368 a_0$, corresponding to a volume per H$_3$S of 110 a$_0^3$. The calculations are performed in a 2$\times$2$\times$2 cubic supercell at the Balderschi k-point, determined at the DFT-PBE level. The DMC nodes are either the LDA ones (first row), or the ones obtained by optimizing the linear coefficients of the one-body Slater orbitals at the VMC level (second row), or the ones yielded by the VMC optimization of both linear coefficients and Gaussian exponents (third row). The optimization is done by VMC energy minimization.
  }
\label{tab:relaxed_det}
\end{table}

As one can see from Tab.~\ref{tab:relaxed_det}, the energy differences, giving an estimate of the barrier height at the DMC level with various nodal surfaces, are all compatible within the statistical error bar of 10 meV/H$_3$S. This represents a relative error smaller than 4\%, given the value of the barrier. Therefore, the fixed node approximation based on the LDA nodes, used throughout the paper, is fully satisfactory, at least for estimating the barriers of the potential energy surface (PES) at a fixed volume and with an accuracy as high as 8 meV/H$_3$S.

\begin{figure}[!h]
\centering
\includegraphics[width=0.6\textwidth]{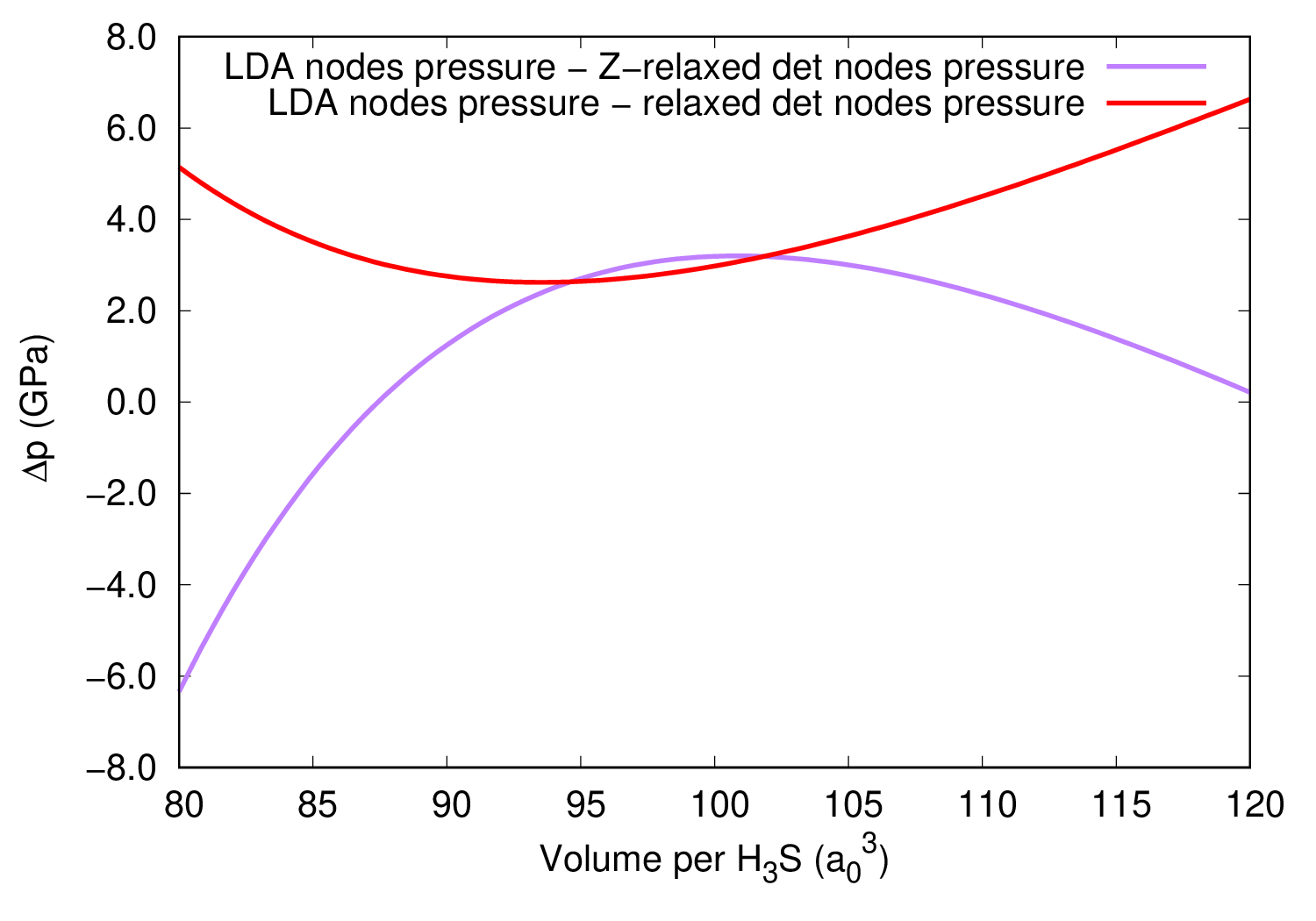}
\caption{
Pressure calibration relation $\Delta p=\Delta p(V)$, determined by taking the EOS based on DMC total energies with LDA nodes as reference. The difference is computed for EOSs obtained by fitting the Vinet model on improved DMC energies as yielded by optimized nodes. Two enhanced nodes are used: the first ones given by the VMC optimization of the linear coefficients of the Slater orbitals (``relaxed det nodes''), the other ones given by the VMC optimization of the full set of variational parameters, comprising both linear coefficients and Gaussian exponents of the Slater orbitals (``Z-relaxed det nodes'').
\label{fig:pofV_poly_diff}}
\end{figure}

Let us study now the impact of the fixed node approximation with LDA nodes to the equation of state (EOS), as determined in the main paper and shown in Fig.~9. As for the barrier height, we carried out a full determinant relaxation, i.e. the optimization of both linear coefficients and the Gaussian exponents of the Slater one-body orbitals, for several volumes. This is the most severe test, because errors are more hardly compensated across different volumes. The optimizations, i.e. VMC energy minimizations, are always performed at the Baldereschi k-point and in the 2$\times$2$\times$2 cubic supercell. Once the nodes optimized, we carried out DMC simulations for different volumes, and then fitted our DMC total energies with a Vinet EOS. The resulting pressure calibration relation, $\Delta p=\Delta p(V)$, is plotted in Fig.~\ref{fig:pofV_poly_diff}, by using the $p=p(V)$ yielded by DMC energies with LDA nodes as reference.

Fig.~\ref{fig:pofV_poly_diff} shows a pressure calibration that, for the best correction, namely for the ``Z-relaxed determinant nodes'', does not exceed a 3.5 GPa difference with respect to the original LDA-nodes EOS in a wide volume range, going from 85 a$_0^3$ to 120 a$_0^3$, thus covering all the most interesting regions of the phase diagram reported in Figs.~8 and 9 of the main paper. Therefore, for the EOS as well as for the PES barriers, the bias due to the DMC fixed node approximation with LDA nodes is negligible. The LDA nodes are accurate enough for high-pressure hydrogen-based systems, as also verified in another system, the high-pressure pristine hydrogen in Ref.~\cite{monacelli2023quantum}.

\section{Convergence of PIMD simulations with respect to the number of beads}

\begin{figure}[!hb]
\centering
\includegraphics[width=0.6\textwidth]{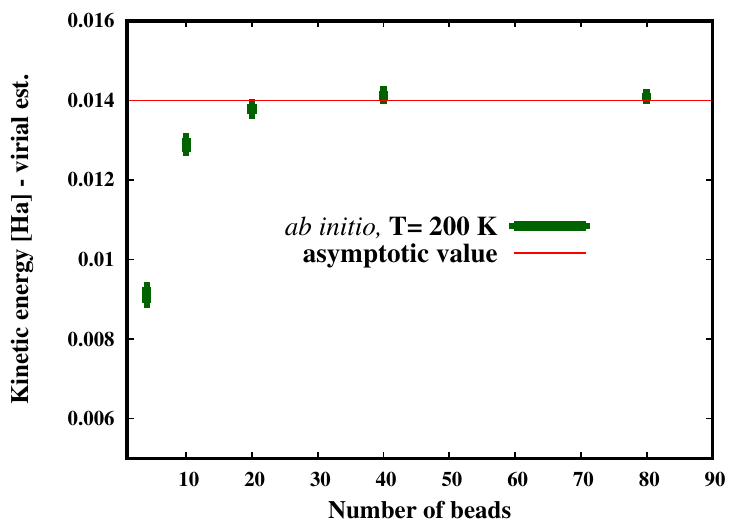}
\caption{Convergence of the kinetic energy virial estimator for V=110 $a_0^3$ from \emph{ab initio} PIMD simulations with respect to the number of beads. \label{fig:convergence_abinitio}}
\end{figure}

In this Section, we present the convergence analysis of the PIMD simulations with respect to the number of beads, both for the \emph{ab initio} and the 3D model simulations using the DFT-BLYP parametrization of the potential energy surface. In Fig. \ref{fig:convergence_abinitio}, we illustrate the convergence of kinetic energy using the virial estimator \cite{Mouhat_2017} from \emph{ab initio} simulations for the volume $V = 110$ a$_0^3$. Notably, with 20 beads, the kinetic energy converges to its asymptotic value (indicated by the red line).

\begin{figure}[!hbt]
\centering
\includegraphics[width=0.6\textwidth]{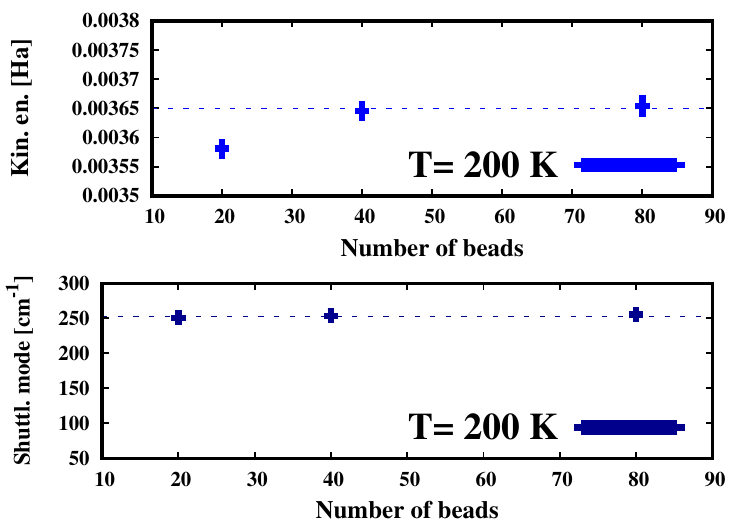}
\caption{Top panel: Convergence of the kinetic energy virial estimator with respect to the number of beads for V=110 $a_0^3$ from simulations of the 3D model using the DFT-BLYP parametrization of the potential. Dashed lines correspond to the asymptotic values. Bottom panel: Convergence of the shuttling mode.  \label{fig:convergence_model}}
\end{figure}

In Fig.~\ref{fig:convergence_model}, we show instead the behaviour of the kinetic energy (top panel of \ref{fig:convergence_model}) and the shuttling mode (bottom panel of \ref{fig:convergence_model}) using the 3D model with the DFT-BLYP parametrization of the PES for V=110 a$_0^3$.
Based on this analysis, we chose
40 beads for our 3D-model PIMD simulations.

\newpage

\section{SSCHA simulations}

\begin{figure}[hb!]
    \centering
    \includegraphics[width=0.49\linewidth]{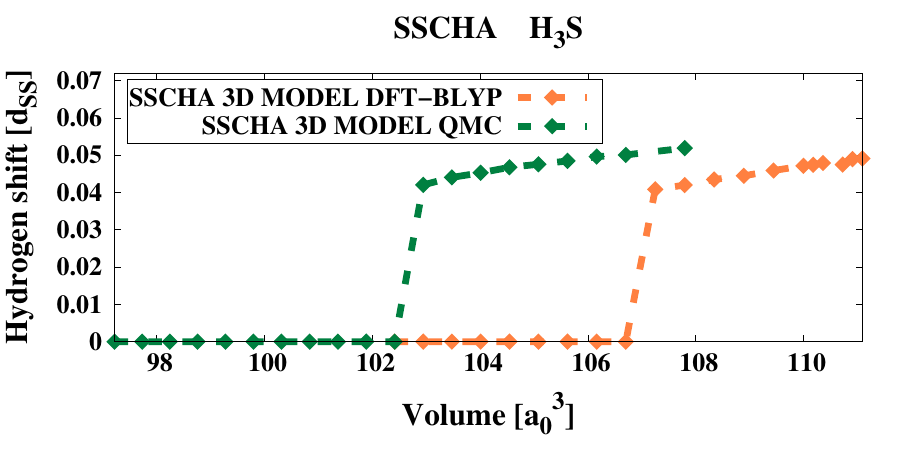}
    \includegraphics[width=0.49\linewidth]{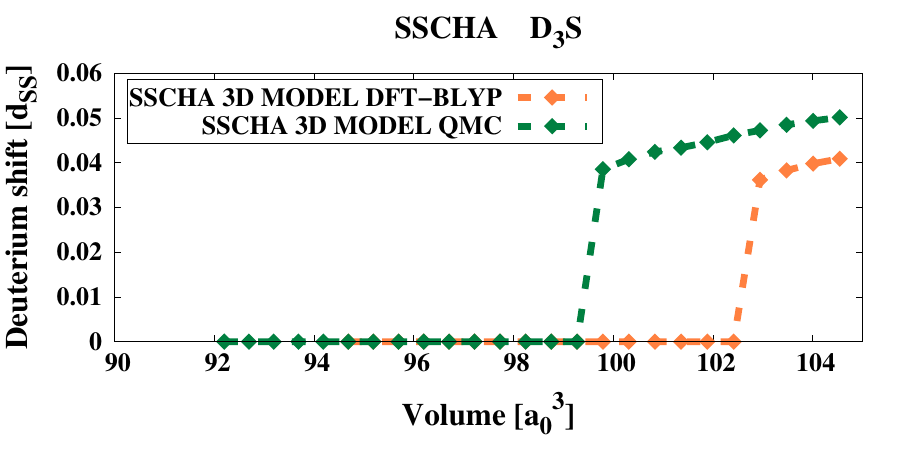}
    \caption{The SSCHA centroid shifted positions of hydrogen atoms from the S-S midpoint are plotted against volume. The left side and right panels show the results for $\ce{H_3S}$ and $\ce{D_3S}$, respectively, using the DFT-BLYP (orange) and QMC (green) 3D-PES. The ferroelectric phase transition is estimated by identifying the jump in positions.}
    \label{fig:susceptibility_SSCHA_SI}
\end{figure}

In this Section, we present the estimation of the ferroelectric phase transition within the SSCHA description of quantum nuclei for the 3D model. In Fig.~\ref{fig:susceptibility_SSCHA_SI}, we illustrate the volume dependence of the centroid positions with the DFT-BLYP and QMC 3D-PES for both $\ce{H_3S}$ and $\ce{D_3S}$. The transition occurs at the volume where the SSCHA centroid position for hydrogen (deuterium) atoms leaves the S-S midpoint. A direct assessment of the isotope effect within the SSCHA framework is obtained by comparing the left and right panels of Fig.~\ref{fig:susceptibility_SSCHA_SI}.

\section{Classical \emph{ab initio} BLYP simulations at 200K}
\label{SI:classic}

\begin{figure}[b!]
    \centering
    \includegraphics[width=0.4\linewidth]{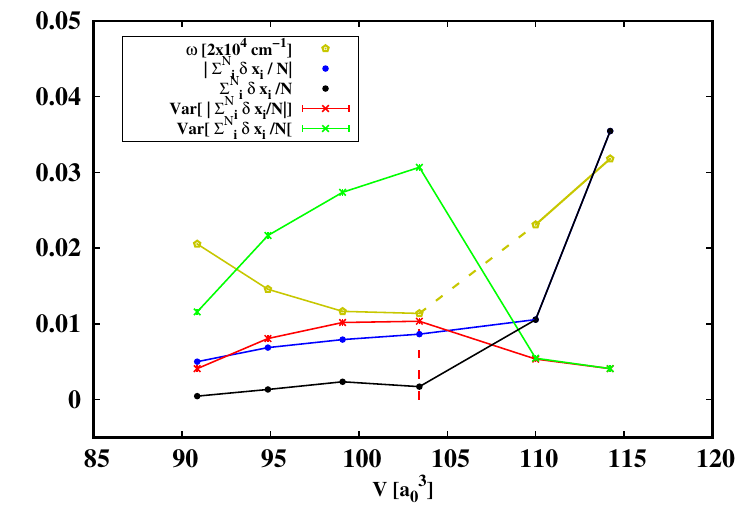}
    \caption{Ferroelectric transition observed through \emph{ab initio} BLYP classical MD simulations at 200K. The golden pentagons illustrate shuttling mode frequencies as a function of volume. The black and blue points represent the ferroelectric order parameter, computed in two ways. The former is derived by averaging the displacements ($\delta x$) of hydrogen atoms concerning the S-S midpoint along MD trajectories, while the latter involves averaging their absolute values. The variances of these order parameters are denoted by green and red crosses, respectively. The red vertical dashed line indicates the peak of the variance, pinpointing the location of the ferroelectric transition.}
    \label{fig:chi_abinitio_classic}
\end{figure}

The ferroelectric transition can be detected by examining the displacements of all $N$ hydrogen atoms with respect to their relative S-S midpoint at a given configurations, and then averaging over a statistical sample of $N$-atom configurations. The corresponding order parameter is defined as $\Delta = \langle \sum_{i=1}^N \delta x_i \rangle$, where the brackets indicate the average over the classical Boltzmann distribution, generated by Langevin molecular dynamics. An analogous order parameter is $\Delta_\textrm{abs} = \langle \lvert \sum_{i=1}^N \delta x_i \rvert \rangle$. In Fig.~\ref{fig:chi_abinitio_classic}, we present the volume dependence of these order parameters, averaged over classical \emph{ab initio} molecular dynamics (MD) trajectories at 200K. We observe a strong correlation between the volume at which hydrogen atoms deviate from the S-S midpoint, the peak of the variance of the order parameters, and the jump in the shuttling mode frequencies.
The remarkable consistency between the values for the critical pressure obtained through different probes underscores the reliability of our results.

\twocolumngrid

\setcounter{page}{1}
\renewcommand{\thepage}{R\arabic{page}} 

\bibliography{sn-bibliography}

\end{document}